\documentclass[12pt,preprint]{emulateapj}

\usepackage{natbib}
\usepackage{graphicx}

\defcitealias{wan15}{Paper~I}

\shorttitle{Effect of transport coefficients on excitation of standing slow-mode waves}
\shortauthors{Wang et al.}

\begin{document}

\title{Effect of transport coefficients on excitation of flare-induced standing slow-mode waves in coronal loops}

\email{tongjiang.wang@nasa.gov}
\author{Tongjiang Wang}
\affil{Department of Physics, Catholic University of America,\\
   620 Michigan Avenue NE, Washington, DC 20064, USA; tongjiang.wang@nasa.gov}
\affil{NASA Goddard Space Flight Center, Code 671, \\
   Greenbelt, MD 20770, USA}

\author{Leon Ofman}
\affiliation{Department of Physics, Catholic University of America,\\
   620 Michigan Avenue NE, Washington, DC 20064, USA}
\affiliation{NASA Goddard Space Flight Center, Code 671, \\
   Greenbelt, MD 20770, USA}
\affiliation{Visiting, Tel Aviv University, Israel}

\author{Xudong Sun}
\affiliation{Institute for Astronomy, University of Hawaii at Manoa,\\
 Pukalani, HI 96768-8288, USA}

\author{Sami K. Solanki}
\affiliation{Max-Planck-Institut f\"{u}r Sonnensystemforschung, \\
Justus-von-Liebig-Weg 3, 37077, G\"{o}ttingen, Germany}
\affiliation{School of Space Research, Kyung Hee University,\\
Yongin, Gyeonggi-Do,446-701, Republic of Korea}

\author{Joseph M. Davila}
\affiliation{NASA Goddard Space Flight Center, Code 671, \\
   Greenbelt, MD 20770, USA}

\begin{abstract}
Standing slow-mode waves have been recently observed in flaring loops by the Atmospheric Imaging Assembly (AIA) of the {\it Solar Dynamics Observatory} ({\it SDO}). By means of the coronal seismology technique transport coefficients in hot ($\sim$10 MK) plasma were determined by \citet[][Paper I]{wan15}, revealing that thermal conductivity is nearly suppressed and compressive viscosity is enhanced by more than an order of magnitude. In this study we use 1D nonlinear MHD simulations to validate the predicted results from the linear theory and investigate the standing slow-mode wave excitation mechanism. We first explore the wave trigger based on the magnetic field extrapolation and flare emission features. Using a flow pulse driven at one footpoint we simulate the wave excitation in two types of loop models: model 1 with the classical transport coefficients and model 2 with the seismology-determined transport coefficients. We find that model 2 can form the standing wave pattern (within about one period) from initial propagating disturbances much faster than model 1, in better agreement with the observations. Simulations of the harmonic waves and the Fourier decomposition analysis show that the scaling law between damping time ($\tau$) and wave period ($P$) follows $\tau\propto{P^2}$ in model 2, while $\tau\propto{P}$ in model 1. This indicates that the largely enhanced viscosity efficiently increases the dissipation of higher harmonic components, favoring the quick formation of the fundamental standing mode. Our study suggests that observational constraints on the transport coefficients are important in understanding both, the wave excitation and damping mechanisms.

\end{abstract}


\section{Introduction}

Fundamental standing slow-mode waves in flaring coronal loops were first discovered with the Solar Ultraviolet Measurements of Emitted Radiation (SUMER) spectrometer onboard {\it Solar and Heliospheric Observatory} ({\it SOHO}) \citep[see a review by][]{wan11}. These oscillations have a period in the range 7--31 minutes and an exponential decay time comparable to the period \citep{wan03a}. The wave modes are identified based on their phase speed, which is close to the speed of sound at the loop's temperature, and a quarter-period phase shift existing between velocity and intensity oscillations \citep{wan02, wan03a, wan03b, yuan15}. Thermal conduction is believed to be the dominant wave damping mechanism \citep{ofm02, dem03}, however, other physical processes such as compressive viscosity, radiative cooling, and heating function may also importantly affect the wave damping in some special conditions \citep[e.g.,][]{alg14, wan15, kum16, nak17}. The slow-mode waves have been applied to derive the magnetic field strength in coronal loops using seismology techniques \citep{wan07, jes16}.

The excitation mechanism of standing slow-mode waves is still poorly understood despite the investment of much effort in both observation and theory. Observations from {\it SOHO}/SUMER showed that excitation of the slow-mode standing waves in hot coronal loops has the following features \citep{wan03a, wan03b, wan05, wan07}: (1) The wave events are often triggered by small (or micro-) flares at one footpoint of the loop with a heating time less than half a wave period. (2) The standing wave patterns are quickly produced within about one wave period (or after only one reflection of the initial disturbance). (3) The loop plasma is impulsively heated to above 6--10 MK, and then cools down gradually. Theoretical analysis and simulations based on 1D loop models show that a footpoint heating pulse with much shorter duration than the wave period generates only (reflected) propagating waves in the loop \citep{tar05, sel05}. \citet{fan15} confirmed this conclusion using a 2.5D MHD model with the similar driver. \citet{sel09} and \citet{ofm12} found that a fundamental standing slow mode wave can be excited quickly in isothermal 3D MHD simulations of hot loops by a fast-mode wave, velocity pulse, or impulsive onset of flows at one footpoint.

\begin{figure*}
\epsscale{1.0}
\plotone{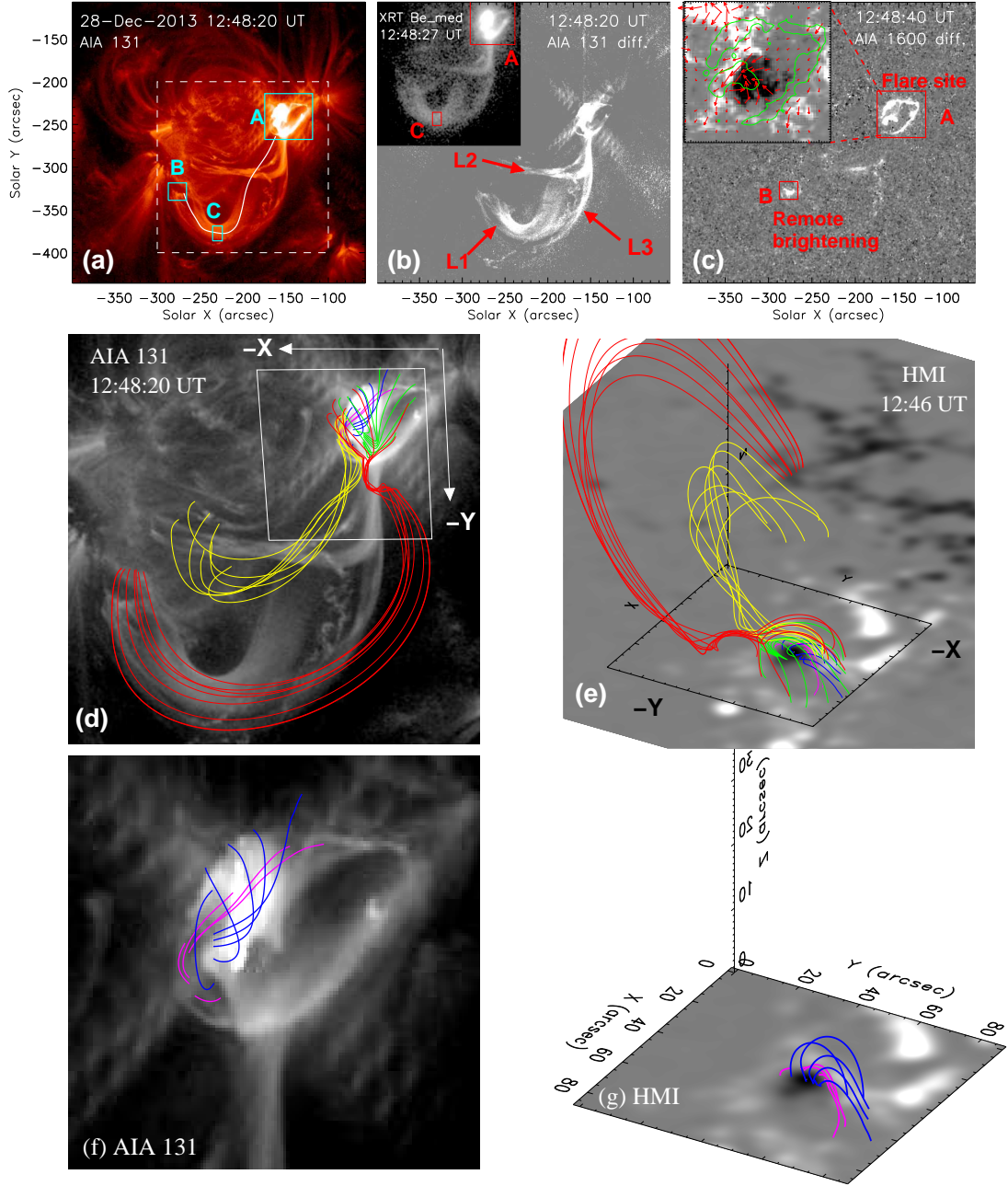}
\caption{ \label{fgflr} Trigger of standing slow-mode waves by a flare in AR 11936 on 2013 December 28. (a) SDO/AIA 131 \AA\ image. The oscillating hot loop (indicated with L1 in (b)) is outlined with a thick white curve. Box A marks the flare region. (b) AIA 131 \AA\ base difference image. L2 and L3 indicate two hot loops that are commonly rooted with loop L1 at the flare site. The inset shows a co-temporal {\it Hinode}/XRT image with the Be-med filter. (c) AIA 1600 \AA\ base difference image. Box B marks a remote brightening at the footpoint of Loop L1. The inset shows contours (at 200 DN~s$^{-1}$) of the flare ribbons in region A overplotted on an HMI vector magnetic field map (observed at 12:46 UT). The background indicates the longitudinal component scaled between $\pm$ 500 G with the positive/negative polarities in white/black colors. The arrows indicate the transverse component with field strength in a range 100--1000 G. The reference AIA images used in (b) and (c) were observed at 12:40 UT prior to the flare. Light curves measured in regions A, B, and C are shown in Figure~\ref{fglcv}. (d) Top view of the magnetic skeleton, superposed on the AIA 131 \AA\ image. The field of view is shown in panel (a) with a dashed box. Field lines (in red, yellow, and green) traced from around a null outline a dome-shaped fan surface and the spine. The pink and blue field lines inside the fan dome show strong shear above the polarity inversion line (PIL). Magnetic reconnection near the null between two flux systems inside and outside the fan may trigger the flare and excite longitudinal waves in the large spine loop. (e) Side view of the magnetic skeleton, superposed on a HMI radial field map (with smoothing and scaled between $\pm$ 1100 G). (f) and (g): Close-up of panels (d) and (e), showing the low-lying, sheared field lines (pink and blue) inside the fan dome.}
\end{figure*}

Recently, the {\it Solar Dynamics Observatory} \citep[$SDO$;][]{pes12}/Atmospheric Imaging Assembly \citep[AIA;][]{lem12} also detected flare-excited longitudinal loop oscillations \citep{kum13, kum15, wan15, man16, nis17}, which bear physical properties similar to the slow-mode waves previously detected with {\it SOHO}/SUMER. From the nearly in-phase temporal relationship between temperature and density disturbances, \citet[][thereafter, Paper I]{wan15} derived that thermal conductivity is strongly suppressed in a flare-heated loop. This result also suggests that compressive viscosity needs to be greatly enhanced to interpret the observed strong wave damping. In Section~\ref{sctobs} we describe and analyze observations of this event to constrain the wave driver and also provide motivations for our modeling study. In Section~\ref{sctmdl} we describe the 1D loop models for simulations of wave excitation. We compare the numerical results of the two types of models with the classical and observationally-constrained transport coefficients in Section~\ref{sctcom}, and analyze the corresponding dissipation properties in Section~\ref{scteff}. We discuss wave trigger and excitation mechanisms as well as the validity of linear theory for small amplitude waves in Section~\ref{sctdis}, and finally present our conclusions in Section~\ref{sctccs}.

\section{Observations}
\label{sctobs}
\subsection{Hints for the wave trigger and loop heating}
\label{scthnt}
A wave event occurring on 2013 December 28 in NOAA Active Region (AR) 11936 was first studied in \citetalias{wan15}, where the transport coefficients were determined from measurements of the wave and plasma thermal properties by coronal seismology. Here we first analyze the event trigger to constrain the wave driver for modeling. Figures~\ref{fgflr}(a) and (b) show that the observed longitudinal waves propagate along a large hot loop (marked L1) seen in the AIA 131 \AA\ band (dominated by Fe\,{\sc xxi}, formed at $\sim$11 MK) and the Be\_med filter of the {\it Hinode}/X-Ray Telescope \citep[XRT;][]{gol07}. The waves were generated by a GOES C3.0-class flare at the footpoint of the loop, which is characterized by circular-like ribbons (see Figure~\ref{fgflr}(c)). We reconstruct the coronal fields using a nonlinear force-free field (NLFFF) extrapolation \citep{wie04, wie06}, based on the photospheric vector magnetic fields observed with {\it SDO}/Helioseismic and Magnetic Imager \citep[HMI;][]{sch12} at 12:46 UT. The magnetic skeleton involving the flare ribbons is calculated (see Figures~\ref{fgflr}(d) and \ref{fgflr}(e)), showing a dome-shaped fan-spine topology. Inside its fan dome the field lines (in pink and blue) overlying the polarity inversion line are strongly sheared (see Panels (f) and (g) for the close-up). The destabilization of the shear field can drive slipping-type reconnection within the fan dome and null-point reconnection, energizing the larger-scale spine loops and eventually powering a flare \citep[e.g.,][]{aul00, mas09, wanh12, sun13}. 

We note that three longer hot loops (marked L1--L3 in Figure~\ref{fgflr}(b)) are associated with the flare, which cannot be well reconstructed by the NLFFF model (see the extrapolated field lines in red and yellow). There are many possible reasons for the mismatching such as non-force-free magnetic configuration of the flaring loops, noise in the boundary condition, non-negligible plasma-beta, imperfect numerical algorithm, etc \citep[see][]{sch08, der15}. Nonetheless, the magnetic topology calculated by the NLFFF extrapolations appears to be basically coincident with the observed emission features, particularly in the vicinity of the footpoints of these loops (e.g., at regions A and B). We suggest that this flare is triggered by slipping-type reconnections at a coronal null point in the fan-spine magnetic topology. The impulsive magnetic energy release heats the large spine loop and the associated pressure disturbances propagate and are reflected back and forth in the hot loop, ultimately forming the standing slow-mode waves. 

\begin{figure*}[t]
\epsscale{1.0}
\plotone{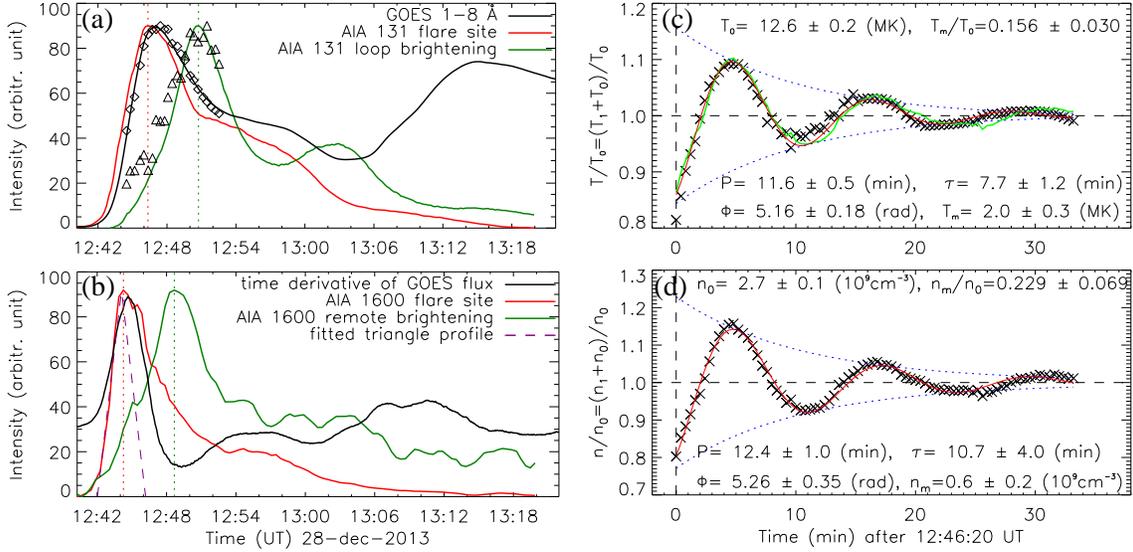}
\caption{ \label{fglcv} (a) Light curves of the flare ribbons (in region A) and the loop brightening (in region C) from AIA 131 \AA\ and XRT Be-med images, and light curve of GOES 1--8 \AA\ soft X-rays (SXRs). Diamonds and triangles show the XRT SXR flux measured from regions A and C, normalized to the maximum of the 131 \AA\ band for comparison. The peak times of light curves in AIA 131 \AA\ band are indicated by vertical dotted lines. (b) Light curves of the flare ribbons (in region A) and the remote brightening (in region B) from AIA 1600 \AA\ images, and time derivative of the GOES SXR flux. The peak times in AIA 1600 \AA\ band are indicated by vertical dotted lines. The impulsive rise phase of the 1600 \AA\ light curve of the flare is fitted to a triangle function (dashed line). (c) Time profile of the temperature (crosses) normalized to the slowly-varying trend measured at region C and the best fit to an exponentially damped sine function (red solid line). (d) Same as (c) but for electron density. The measured physical parameters of the waves are marked on the plots. The green solid curve in (c) is the predicted variation of temperature derived from the observed density variation ($n/n_0$) for an adiabatic process. Panels (c) and (d) are from \citet{wan16}.}
\end{figure*}

To explore the timing between loop heating and wave excitation, we compare light curves of the EUV/UV and soft X-ray (SXR) emissions measured at different locations. Figure~\ref{fglcv}(a) shows that the flare emissions measured at region A in the AIA 131 \AA\ band and XRT filter evolve coincidentally with the GOES SXR flux in the rise phase. They peaked almost simultaneously at about 02:46 UT, preceding the peak time of the loop brightening measured at region C by $t_{\rm AC}$=264 seconds. If we assume that the loop brightening at region C is caused by the initial (compression) disturbance traveling (or the injected hot plasma moving) from the flare site A, its propagation speed can be estimated from the travelled distance ($L_{\rm AC}$=110 Mm) along the 3D loop \citep[see][]{wan16}. We obtain $V_p=L_{\rm AC}/t_{\rm AC}$=417 km~s$^{-1}$, which corresponds to the sound speed ($C_s=(\gamma{k_B}T/\mu{m_p})^{1/2}=166(T/{\rm MK})^{1/2}$ km~s$^{-1}$) at temperature $T$=6.3 MK, where $\gamma$=5/3, $k_B$ is the Boltzmann constant, and $m_p$ the proton mass. We take $\mu$=0.5 to be consistent with that used for simulations in Section~\ref{sctmdl}.

We use the AIA UV 1600 \AA\ light curve (flare emission dominated by C\,{\sc iv}, from the upper chromosphere to transition region) to characterize the heating source in flares as the 1600 \AA\ emission like the hard X-ray (HXR) emission indicates the immediate response of the lower atmosphere to impulsive energy deposit \citep{fish85, qiu12, liu13}. Figure~\ref{fglcv}(b) shows that the peak time (02:44 UT) of flare emission at 1600 \AA\ is consistent with that of the GOES flux time derivative, which is a proxy of the HXR light curve during the impulsive phase \citep{den93}. Using the method of \citet{qiu12}, we assume the heating function to be symmetric in time (here taken as a full triangle) and estimate the heating duration ($t_{\rm dur}$) as twice the rise time of the impulsive phase in 1600 \AA. We obtain $t_{\rm dur}$=4.2 minutes based on linear fitting (see the dashed line in Figure~\ref{fglcv}(b)). We will use this measurement to constrain the duration of the wave driver in simulations.

Heating of the spine loop is also evidenced by an occurrence of brightening at its remote footpoint (called the remote brightening relative to the flare site) seen in the AIA 1600 \AA\ band (marked B in Figure~\ref{fgflr}(c)). The remote brightening may be caused by energized particles or intense heat flux flowing from reconnections near the null along the spine loop \citep{mas09, sun13}. Figure~\ref{fglcv}(b) shows that the 1600 \AA\ remote brightening is delayed by 264 seconds in peak time compared to the 1600 \AA\ flare emission. It is plausible to assume that the loop here is heated by a thermal front because heating by energetic nonthermal particles typically happens on a much shorter timescale \citep{asc04}. For example, for a coronal loop of length $L$=200 Mm, it takes an electron at relativistic speed less than 7 seconds to travel from one end to the other. We may also exclude the slow shock as a possible heating source as the observed waves bear very linear properties (see Figures~\ref{fglcv}(c) and (d)). Taking this time lag to be the traveling time of the heating front along the whole loop, we estimate its propagation speed $V_h$=680 km~s$^{-1}$ from the 3D loop length \citep[$L\simeq$180 Mm;][]{wan16}. We find that the heat propagation is faster than the wave disturbance ($V_h\sim 1.6 V_p$). This suggests that the spine loop L1 may be heated preceding the arrival of the initial disturbance at the remote footpoint. This may explain the fact that the initial disturbance propagates with a speed close to the speed of sound in the plasma of $T\gtrsim$ 6 MK. In addition, some numerical simulations also showed that thermal fronts propagate with a speed faster than sound waves and the evaporated hot flows in hot flare loops \citep{arb09, liu09}.

\subsection{Motivation for modeling}
Based on the finding that the density and temperature oscillations are nearly in phase (see Figures~\ref{fglcv} (c) and (d)), we concluded in \citetalias{wan15} that the thermal conduction is strongly suppressed and the compressive viscosity is the dominant wave damping mechanism. The suppression of thermal conductivity in hot loops implies that variations in temperature and density approximately follow an adiabatic relation, $T/T_0 = (n/n_0 )^{\gamma-1}$, when wave amplitudes are small. Here $\gamma$=5/3, $T$ and $n$ are the temperature and number density of the plasma, $T_0$ and $n_0$ are the corresponding slowly-varying trend. Figure~\ref{fglcv}(c) shows that the predicted temperature variation from the adiabatic process agrees well with the observed data, supporting this supposition. Our first motivation in this article is to validate the seismological results obtained in \citetalias{wan15} based on the linear wave theory, using more advanced 1D nonlinear MHD simulations. We will show that the models with the seismology-determined transport coefficients can reproduce the observed wave properties much better than the models with the transport coefficients calculated from the classical \citep{spi56} theory.  

Our second motivation is to understand how the fundamental standing slow-mode wave can be excited in a very short timescale by a footpoint flare as observed in this event. Figures~\ref{fglcv}(c) and (d) show that the temperature and density variations agree well with a (damped) sinusoidal function with the period that is close to that of the fundamental mode ($P=2L/C_s$=12 minutes). It is peculiar that the observed wave with initial large amplitudes ($V/C_s\approx{n_m/n_0}$=0.23) manifests neither the nonlinear effect (such as the steepened front) nor the coexistence with higher harmonics. We will show that the models with the seismology-determined transport coefficients can successfully generate the fundamental standing mode with excitation time and wave properties consistent with the observation while the models with the classical transport coefficients will fail. Our analysis suggests that the more efficient dissipation of higher harmonic components in initial disturbances due to the large enhancement of viscosity may be the main cause for the quick formation of the fundamental mode.

\section{Loop models}
\label{sctmdl}
To simulate the propagation of slow-mode waves in a coronal loop, we solve the nonlinear one-dimensional MHD equations in Cartesian geometry. The magnetic field of the loop is taken to be along the $x$-direction, and it enters into the model only as a wave guide. The gravity is neglected since 
the loop height is much smaller than the pressure scale height ($H\simeq$500 Mm) for hot plasma of T$\simeq$10 MK. We also neglect radiative losses in the energy equation as the radiation cooling timescale ($\tau_{\rm rad}\simeq$570 minutes) is much longer than the oscillation period (see discussions in \citetalias{wan15}). The equations including the terms for compressive viscosity and thermal conduction are
\begin{eqnarray}
 \frac{{\partial}{\rho}}{\partial{t}} +\frac{\partial}{\partial{x}} (\rho{V})&=&0, \\
 \rho\left(\frac{{\partial}{V}}{\partial{t}}+V\frac{\partial{V}}{\partial{x}} \right)&=&-\frac{{\partial}{p}}{\partial{x}}+ F_{\nu}, \\
 \frac{{\partial}{T}}{\partial{t}}+(\gamma-1)T\frac{\partial{V}}{\partial{x}}+V\frac{\partial{T}}{\partial{x}}&=&\frac{(\gamma-1)m_p}{2k_B}\left(\frac{1}{\rho}\right)(S_{\nu}+H_c).
 \end{eqnarray}
The viscous force due to compressive viscosity is $F_{\nu}=(4/3)\eta_0(\partial^2V/\partial{x}^2)$, the viscous heating term is $S_{\nu}=(4/3)\eta_0(\partial{V}/\partial{x})^2$, and $\gamma$=5/3. The classical {\it Braginskii} compressive viscosity coefficient is given by
$\eta_0=2.23\times10^{-15}T_i^{5/2}/{\rm ln}\Lambda$ g~cm$^{-1}$s$^{-1}$, where $T_i$ is the ion temperature (considering protons only), and ${\rm ln}\Lambda=8.7-{\rm ln}(n^{1/2}T_i^{-3/2})$ is the Coulomb logarithm, weakly dependent on $T_i$ and the number density $n$. The heat conduction term along the magnetic field ($x$-direction) is $H_c=\partial/\partial{x}\,[\kappa_\|(\partial{T_e}/\partial{x})]$, where $T_e$ is the electron temperature and $\kappa_\|$ is the classical {\it Spitzer} thermal conductivity parallel to the magnetic field given by $\kappa_\|=7.8\times10^{-7}T_e^{5/2}$ ergs cm$^{-1}$s$^{-1}$K$^{-1}$ \citep{spi56, spi62}. In the single-fluid MHD model employed here, the electron and ion temperatures are assumed to be equal ($T_e=T_i=T$). We also assume that the loop density and temperature are initially uniform. In the numerical simulations, we use the following loop parameters measured in \citetalias{wan15}: the loop length $L$=180 Mm, the density $n_0=2.6\times10^9$ cm$^{-3}$, and the temperature $T_0$=9 MK. With these parameters, the corresponding sound speed $C_s$=498 km~s$^{-1}$, the classical compressive viscosity is $\eta_0$=24.6 g~cm$^{-1}$s$^{-1}$, and the heat conduction $\kappa_\|$=1.90$\times10^{11}$ ergs~cm$^{-1}$s$^{-1}$K$^{-1}$.

To simulate the flare-induced perturbation, we inject an impulsive flow along the magnetic field at the boundary,
\begin{equation}
V(x=0, t)  =   \left\{
\begin{array}{ll}
\frac{1}{2}V_0 \left[1-{\rm cos}\left(\frac{2\pi t}{t_{\rm dur}}\right)\right] &  \quad (0\leqslant t \leqslant t_{\rm dur}),\\ 
 0 &  \quad (t > t_{\rm dur}). \label{eqdrv}
\end{array}\right..
\end{equation}
We take the pulse amplitude $V_0=(n_m/n_0)C_s$=0.23\,$C_s$=115 km~s$^{-1}$, where $n_m/n_0$ is the measured maximum amplitude of the density perturbations (see Figure~\ref{fglcv}(d)). We take the pulse duration $t_{\rm dur}$=4 minutes based on the loop heating duration obtained in Section~\ref{scthnt}. The MHD equations are solved by adopting the fourth-order Runge-Kutta method in time and fourth-order derivatives in space using 256 grid points \citep{ofm02}. Numerical convergence is tested by doubling the resolution and comparing the results. The boundary conditions at both ends of the loop are $V(0, t)$=$V(L, t)$=0 (except the flow injection $V(0, 0\leqslant t\leqslant t_{\rm dur}$)) and zero-order extrapolation for the rest of the variables.

\begin{figure*}
\epsscale{1.0}
\plotone{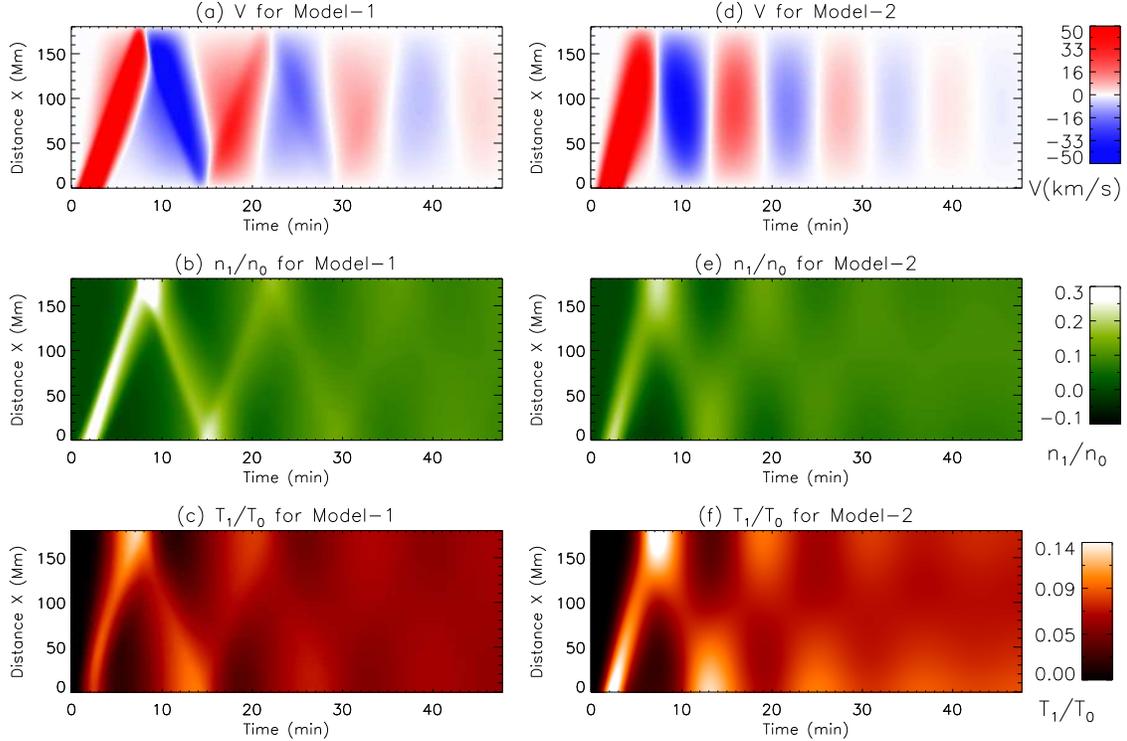}
\caption{ \label{fgsim} Comparison between two models for slow-mode wave excitation by a flow pulse at footpoint $x=0$ Mm of the loop with length $L=180$ Mm. (a)$-$(c) Time distance maps for velocity ($V$), perturbed density ($n_1/n_0$), and perturbed temperature ($T_1/T_0$) along the loop simulated based on the model with the classical thermal conduction and classical compressive viscosity (Model 1). (d)$-$(f) Same as (a)$-$(c) but based on the model with the observation-constrained transport coefficients, {\it i.e.} the zero-value conductivity and 15-times enhanced viscosity (Model 2). }
\end{figure*}

\begin{deluxetable*}{ccrrcl}
 \tabletypesize{\scriptsize}
 \tablecaption{Propagation Speed Measurements of Initial Disturbances in Various Cases of Simulations 
 \label{tabpsp}}
 \tablewidth{0pt}
 \tablehead{ 
\colhead{Case} & \colhead{Conductivity} & \colhead{Viscosity Coeff.} &\colhead{Flow Amplitude} & \colhead{Propagation Speed} & \colhead{Model}
}
\startdata
1 & $\kappa_\|^{\rm class}$ & $\eta_0^{\rm class}$ & 0.01 $V_0$ & 0.81 $C_s$ & Model 1A \\
2 & $\kappa_\|^{\rm class}$ & $\eta_0^{\rm class}$ &      $V_0$ & 1.01 $C_s$ & Model 1\\
3 & 0 & 15 $\eta_0^{\rm class}$                    & 0.01 $V_0$ & 1.17 $C_s$ & Model 2A\\
4 & 0 & 15 $\eta_0^{\rm class}$                    &      $V_0$ & 1.24 $C_s$ & Model 2\\
\hline
5 & 0 & 0                                          & 0.01 $V_0$ & 1.00 $C_s$ & -- \\
6 & 0 & 0                                          &      $V_0$ & 1.31 $C_s$ & -- \\
7 & $\kappa_\|^{\rm class}$ & 0                    & 0.01 $V_0$ & 0.79 $C_s$ & -- \\
8 & $\kappa_\|^{\rm class}$ & 0                    &      $V_0$ & 1.09 $C_s$ & -- \\
9  & 0 & $\eta_0^{\rm class}$                      & 0.01 $V_0$ & 1.00 $C_s$ & -- \\
10 & 0 & $\eta_0^{\rm class}$                      &      $V_0$ & 1.27 $C_s$ & --\\
\enddata
\tablecomments{Column 1 is the case number. Column 2 is the thermal conductivity parallel to the magnetic field. Column 3 is the compressive viscosity coefficient. Column 4 is the amplitude of a flow pulse with $V_0$=0.23 $C_s$, where $C_s$=498 km~s$^{-1}$ is the adiabatic sound speed at $T$=9 MK. Column 5 is the measured propagation speed for initial disturbances. Column 6 gives the name of the model whose simulations are shown in the paper. 
 }
\end{deluxetable*}

\section{Comparison of simulations between the two models}
\label{sctcom}

Using the 1D MHD model described above, we simulate the flare-generated standing slow-mode waves reported in \citetalias{wan15} for two cases: the first model with the classical transport coefficients $\kappa_\|$=1.90$\times10^{11}$ ergs~cm$^{-1}$s$^{-1}$K$^{-1}$ and $\eta_0$=24.6 g~cm$^{-1}$s$^{-1}$ (thereafter, called Model 1), and the second model with the observation-constrained transport coefficients: $\eta_0^{\rm obs}$=15 $\eta_0$ when assuming $\kappa_{\|}^{\rm obs}$=0 obtained in \citetalias{wan15} (Model 2).  Figure~\ref{fgsim} compares the temporal evolution of velocities ($V$), perturbed densities ($n_1/n_0\equiv(n-n_0)/n_0$), and perturbed temperatures ($T_1/T_0\equiv(T-T_0)/T_0$) along the loop between the two models. Here $T$ and $n$ are the temperature and number density of the plasma, $T_0$ and $n_0$ are the corresponding equilibrium quantities. The `zigzag' pattern, which is obvious in the first two wave periods for $V$ and $T_1/T_0$, suggests that a propagating wave is excited and undergoes reflections from the footpoints in Model 1. The propagating wave tends to transition to the standing wave after 4--5 reflections as indicated by the formation of in-phase oscillations along the loop in $V$. Whereas in Model 2 a fundamental standing wave is excited immediately after the reflection of the initial perturbations at the remote footpoint ($x=L$) as evidenced by the spatial and temporal features: (1) the velocity perturbations along the loop are in phase, (2) the two legs of the loop oscillate in antiphase in $n_1/n_0$ and $T_1/T_0$, and (3) the oscillations between $V$ and $n_1/n_0$ have a 1/4-period phase shift. In addition, the standing slow-mode wave pattern formed in the simulations is in accord with that predicted by linear theory of a cylinder model \citep[comparing with Figure~3 in][]{yuan15}.

\begin{figure}
\epsscale{1.0}
\plotone{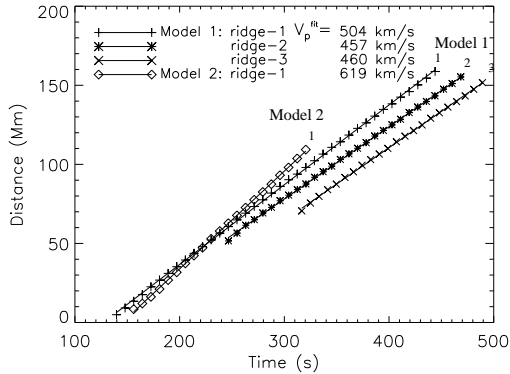}
\caption{ \label{fgvp} The peak position of density perturbations along the loop measured for three ridges of Model 1: ridge-1 (pluses) during $t$=[138, 444] seconds, ridge-2 (asterisks) during $t$=[624, 846] seconds, and ridge-3 (crosses) during $t$=[1086, 1260] seconds.  For ridge-2 the peak positions are measured from $x=L$ and are plotted with the time of $t-378$ seconds. For ridge-3 the peak positions are plotted with the time $t-768$ seconds. The peak positions of ridge-1 (diamonds) for Model 2 are measured during $t$=[156, 318] seconds. The solid lines are the best fit to the data points, and their slope values ($V_p^{\rm fit}$) are marked on the plot. }
\end{figure}

We estimate the propagation speeds of density perturbations by measuring the slope of ridges seen in Figures~\ref{fgsim}(b) and (e). Figure~\ref{fgvp} shows the linear fits to the peak positions of the first three ridges of $n_1/n_0$ in Model 1 and the first ridge in Model 2. From the slope of the ridges, we estimate the wave propagation speed $V_p^{\rm fit}$=1.01, 0.92, and 0.92 $C_s$ for Model 1, and  $V_p^{\rm fit}$=1.24 $C_s$ for Model 2, where $C_s$=498 km~s$^{-1}$ is the adiabatic sound speed for the loop at $T_0$. Note that the initial perturbations (before reflection) in Model 2 can be regarded as a propagating wave, so we can estimate its phase speed from the slope of the ridge, but this technique cannot be used to estimate the phase speed for the standing wave that has been established. We find that the waves propagate with a phase speed close to the speed of sound, and the propagation speed of initial perturbations in Model 2 is supersonic and higher than in Model 1 (by $\sim$23\%). 

To investigate the factors that affect the propagation speed of the initial perturbation, we run the simulations for various cases and list the measured phase speeds in Table~\ref{tabpsp}. Through a comparison between the two cases that vary only in one parameter (i.e., the control parameter), we examine its influence on the wave speed. The control parameter could be a different physical quantity in each set of runs. For short, we define the scenario of Case i vs.~Case j as testing how the wave speed depends on the control parameter by comparing the modeling results of the two cases. We find that the propagation speed of the initial perturbation strongly depends on the initial flow amplitude, and the effect is most evident in the scenarios where no viscosity is included. For example, large amplitude pulses increase the phase speed by more than 30\% due to strong nonlinearity that steepens the wave front in the scenarios of Case 5 vs.~Case 6 and Case 7 vs.~Case 8. The perturbation speed also depends on thermal conductivity. The high thermal conduction leads to the waves propagating at the lower, (near-)isothermal sound speed (see the discussion in Section~\ref{sctvwt}), whereas strongly suppressed conduction causes the wave propagation at the higher, adiabatic sound speed (e.g., the scenarios of Case 7 vs.~Case 5 and Case 1 vs.~Case 9). The compressive viscosity plays a weak role in changing the wave propagation speed and its effect is different for small and large flow amplitudes. In the case of a small amplitude pulse, the viscosity enhancement increases the wave speed due to its dispersive effect on the wave (e.g., the scenarios of Case 9 vs.~Case 3 and Case 7 vs.~Case 1). In the case of a large amplitude pulse, the viscosity enhancement slightly reduces the wave speed likely due to its smoothing effect on nonlinearity (e.g., the scenarios of Case 10 vs.~Case 4 and Case 8 vs.~Case 2). The above analysis suggests that the fact that the initial propagation speed of Model 2 is higher than that of Model 1 mainly results from the suppression of thermal conduction (see the scenario of Case 2 vs.~Case 10 vs.~Case 4).

\begin{figure*}
\epsscale{1.0}
\plotone{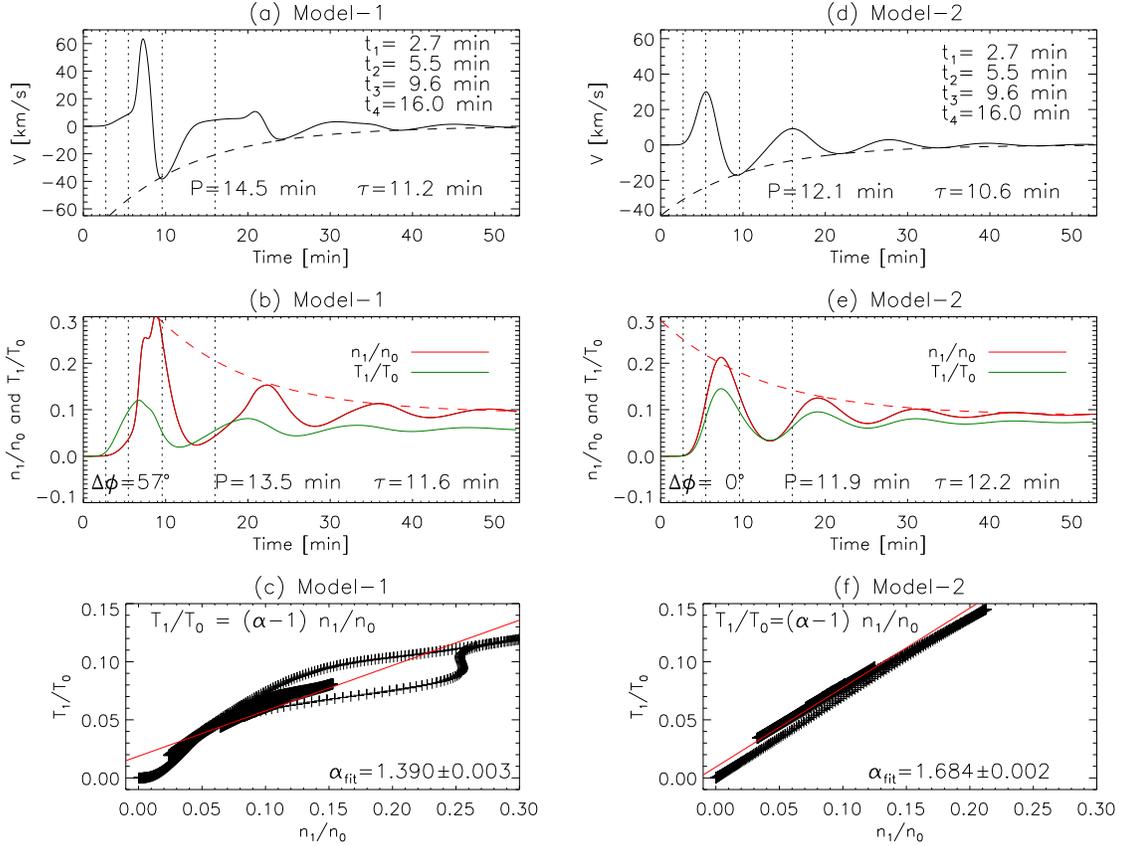} 
\caption{ \label{fgtpf} Temporal evolution of (a) the velocity $V$, and (b) the perturbed density $n_1/n_0$ and temperature $T_1/T_0$ at the location $x=158$ Mm for Model-1. The exponential decay time fit follows the dashed line. (c) The scatter plot of perturbed density and temperature (pluses) and its best fit (solid line) for Model-1. The measured oscillation period ($P$), decay time ($\tau$), phase shift ($\Delta{\phi}$) between $n_1$ and $T_1$, and polytropic index ($\alpha_{\rm fit}$) are marked on the plots. (d)$-$(f) Same as (a)$-$(c) but for Model-2. The vertical dotted lines in the top four panels indicate four times, called  $t_1-t_4$, at which the wave spatial profiles are shown in Figure~\ref{fgspf}.}
\end{figure*}

Figure~\ref{fgtpf} compares the temporal evolution of velocity, density, and temperature perturbations between the two models, measured at a location ($x=0.88 L$) of the loop near the remote footpoint. It indicates that identifying whether the waves are propagating or standing can also be based on their temporal features at a fixed spatial location. The time profiles of $V$ and $n_1/n_0$ for Model 1 clearly deviate from a (damped) sinusoidal function, or they are a non-sinusoidal wave (see Figures~\ref{fgtpf}(a) and (b)), while the time profiles for Model 2 look nearly to be a harmonic wave (see Figures~\ref{fgtpf}(d) and (e)). We estimate the wave period ($P$) by averaging time intervals between successive peaks in the velocity (or density) profile, and estimate the damping time ($\tau$) by fitting the wave peaks to an exponentially-damped function ($f(t)=A_0 + A_1\,{\rm exp}(-t/\tau)$). The measured $P$ and $\tau$ are marked on the plots. We find that the waves simulated by the both models have period and damping time close to the observed values ($P_{\rm obs}$=12.4 minutes and $\tau_{\rm obs}$=10.7 minutes for $n_1/n_0$). It is noted that the wave period of Model 2 is slightly shorter than that of Model 1, consistent with the fact that the phase speed in Model 2 is higher than that in Model 1 as measured above.

\begin{figure}
\epsscale{1.0}
\plotone{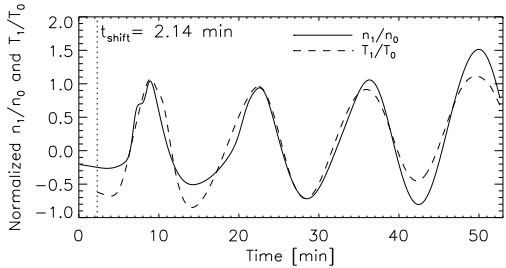} 
\caption{ \label{fgntn} The amplitude-normalized density (solid line) and temperature (dashed line) perturbations obtained from Model 1. The temperature profile has been corrected relative to the density profile by a phase shift of $t_{\rm shift}$=2.14 minutes, which corresponds to the maximum correlation between them. }
\end{figure}

The simulations show the presence of a large phase shift between density and temperature perturbations for Model 1 (see Figures~\ref{fgtpf}(b)) and a nearly in-phase relationship between them for Model 2 (see Figures~\ref{fgtpf}(e)), confirming the predicted results from linear MHD theory (see discussions in Section~\ref{sctvwt}). We measure the phase shift ($\Delta{\phi}$) by applying the cross correlation to the time profiles of $n_1/n_0$ and $T_1/T_0$ which are first normalized to the damped amplitudes by $(s(t)-s_0)/(f_s(t)-s_0)$, where $s$ represents $n_1/n_0$ or $T_1/T_0$, $f_s(t)$ is the best-fit exponentially-damped function, and $s_0$ is the average of $s(t)$ over time. For Model 1 we obtain a time shift $t_{\rm shift}$=2.14 minutes between $n_1/n_0$ and $T_1/T_0$ (see Figure~\ref{fgntn}), and calculate the phase shift as $\Delta{\phi}=360{\degr}(t_{\rm shift}/P)=57{\degr}$, where $P$=13.5 minutes is the wave period measured for $n_1/n_0$. For Model 2 we obtain $\Delta{\phi}=0{\degr}$ using the same method.

Assuming that a polytropic description holds for the loop gas, so $p\propto\rho^\alpha$, where $p$, $\rho$, and $\alpha$ are the gas pressure, mass density, and polytropic index, and applying the ideal equation of state, we obtain $T/T_0=(n/n_0)^{\alpha-1}$. Taking $T=T_0 + T_1$ and $n=n_0 + n_1$, the following relationship can be derived using the linear approximation \citep[e.g.,][]{van11}:
\begin{equation}
  \frac{T_1}{T_0}=(\alpha-1)\frac{n_1}{n_0}.
\end{equation} 
We measure the polytropic index $\alpha$ by fitting the scaling between $T_1/T_0$ and $n_1/n_0$ after first removing their phase shift $\Delta{\phi}$. Using this method we obtain $\alpha=1.390\pm0.003$ for Model 1 (see Figure~\ref{fgtpf}(c)), and $\alpha=1.684\pm0.002$ for Model 2 (see Figure~\ref{fgtpf}(f)). We find that the measured value of $\alpha$ in Model 2 is very close to the adiabatic index $\gamma$=5/3 as measured from the observational data in \citetalias{wan15}.

\begin{figure*}
\epsscale{1.0}
\plotone{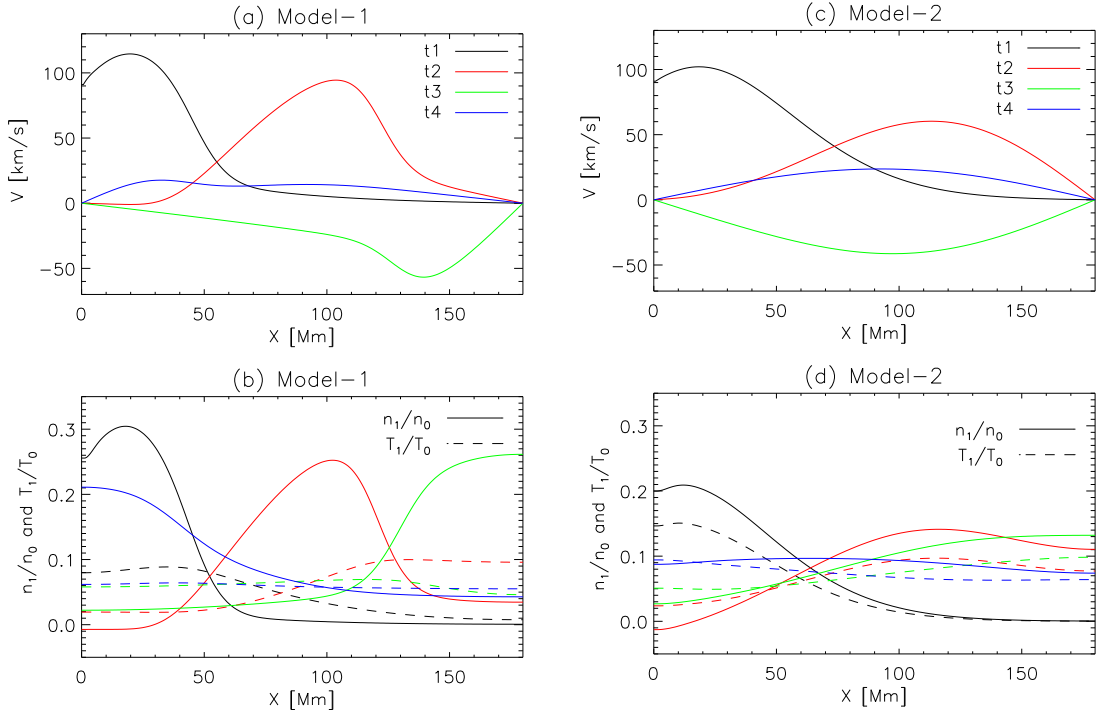}
\caption{ \label{fgspf} Spatial distributions of (a) the velocity, and (b) the perturbed density (solid line) and perturbed temperature (dashed line) along the loop at $t=$ 2.7, 5.5, 9.6, and 16.0 minutes (indicated with $t_1-t_4$ in Figure~\ref{fgtpf}) for Model-1.  (c) and(d) Same as (a) and(b) but for Model-2. The accompanying animation shows the evolution of the velocity, density, and temperature perturbations in the two models from $t$=0.0 minutes to $t$=47.7 minutes. The animation duration is 7 s. (An animation of this figure is available.) }
\end{figure*}

Figure~\ref{fgspf} compares the evolution of the perturbed velocity, density, and temperature profiles along the loop of the two models. The difference is noticeable in the profiles at the times $t_1$ and $t_2$ between Model 1 and Model 2. The velocity and density pulses in Model 2 are much more spread out than those in Model 1. This feature is caused by the significant enhancement of the viscous force $F_{\nu}$ in Model 2, which is 15 times higher than that in Model 1 (see discussions in Section~\ref{sctwem}). The higher viscous force greatly reduces the spatial gradients of velocity or efficiently smooths the velocity pulse in space. The effect is equivalent to effectively increasing the dissipation of higher harmonics in the waves. Animations of Figure~\ref{fgspf} (available in the online version) are also helpful in identifying the mode (propagating or standing) of the excited waves. The animations show that for Model 2 the velocity oscillations become nearly in-phase along the loop (indicating a setup of standing waves) after the propagating pulse reflects once, while for Model 1 it takes many reflections for the propagating pulse to form the in-phase oscillations.  

\section{Dissipation of higher harmonics}
\label{scteff}
The simulations have shown that for the same initial and boundary conditions the standing wave can be set up much quicker in Model 2 than in Model 1. This implies that the change of the transport coefficients from the classical values to the observation-constrained values leads to more efficient dissipation of higher harmonic components in the initial pulse. In this section we analyze the difference in dependence of the damping rate on wave frequency for high harmonics between the two models using two methods.

\begin{figure*}
\epsscale{1.0}
\plotone{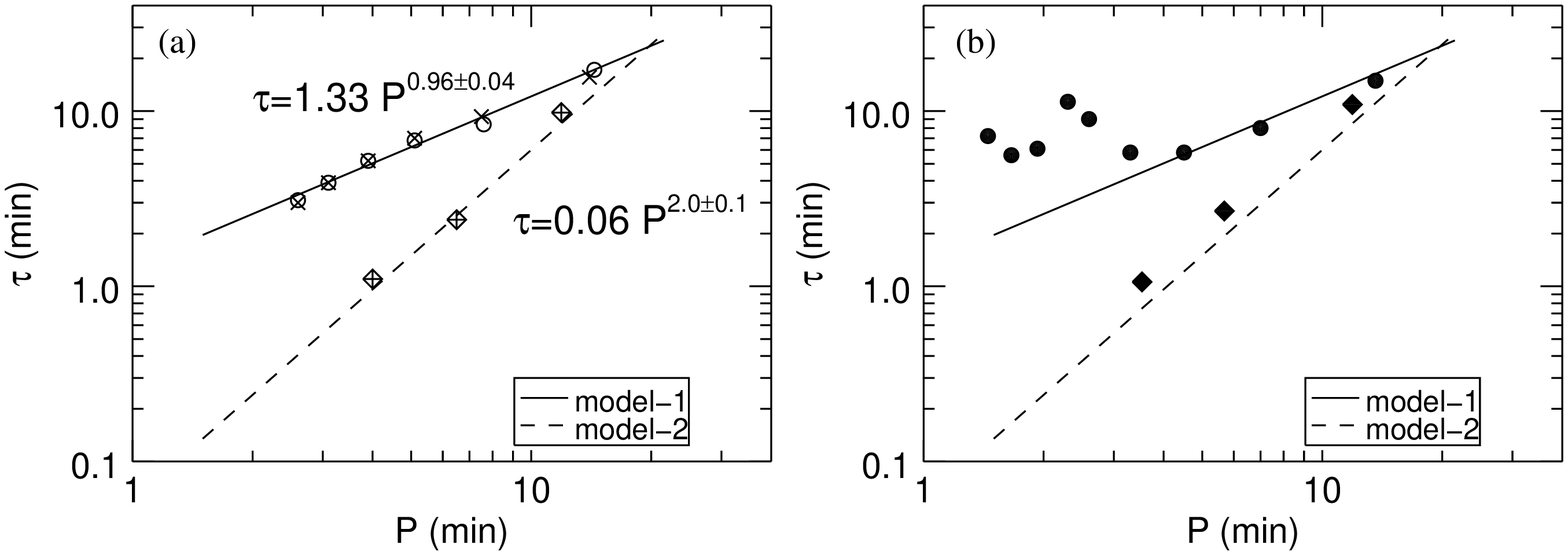}
\caption{ \label{fgtdp} (a) Damping time versus period of the velocity oscillations with different wavenumbers for Model-1 (circles and crosses) and Model-2 (diamonds and pluses). The circle and diamond symbols represent the case with initial velocity amplitude $V_0=0.23\,C_s$, while the cross and plus symbols represent the case with $V_0=0.023\,C_s$. The solid and dashed lines are the best fit power-law functions of the form $\tau=aP^b$. (b) Damping time versus period of the decomposed Fourier components for Model-1 (filled circles) and Model-2 (filled diamonds). The solid and dashed lines are the same as in (a).}
\end{figure*}

In the first method, we simulate the standing waves by setting the initial velocity profile in the form as used in \citet{ofm02}
\begin{equation}
V(x, t=0)=V_0\,{\rm sin}(k\pi{x}/L),
\label{eqvx0}
\end{equation}
where $V_0$ is the amplitude of the wave at $t$=0, and $k$ is the harmonic number with values of 1, 2, 3, $...$ corresponding to the fundamental mode, second harmonic, third harmonic, $...$. The harmonic waves with $k$=1--6 for the two models are simulated with two different initial amplitudes $V_0=0.23\,C_s$ and $0.023\,C_s$. We measure the wave period and damping time from the velocity oscillations using the same method as in Section~\ref{sctcom}. Figure~\ref{fgtdp}(a) shows the damping time with the period and the best-fit scaling for the two models. We find that the power of the scaling is 0.96$\pm$0.04 for Model 1 and 2.0$\pm$0.1 for Model 2. It is noted that each case (in $V_0$) of Model 2 shows only three data points, corresponding to the harmonics $k$=1--3. Because the higher harmonics with $k\geqslant4$ are damped out within one wave period, the measurement of their damping times becomes uncertain. In addition, we find that the measurement results of wave period and damping time in the simulations with $V_0=0.23\,C_s$ and $0.023\,C_s$ are nearly the same, indicating that the obtained scaling laws for the two models are insensitive to the variability in the initial amplitude of different harmonics. The approximate linear scaling ($\tau\propto{P}$) between the damping time and wave period for Model 1 agrees with the result obtained in \citet{ofm02} based on a similar model. The slope of this scaling is smaller than that ($\tau\propto{P^2}$) expected by linear slow wave dissipation theory \citep{por94}. \citet{ofm02} attributed the smaller slope to the nonlinearity of observed oscillations. Here our simulations indicate that the scaling $\tau\propto{P}$ holds also for the waves with small amplitudes ($V_0/C_s$=0.023), suggesting that the small dissipation approximation used in the derivation of the scaling relation by linear theory cannot be met in our case. In the other word, the nonlinear effect (including the scaling $\tau\propto{P}$) for Model 1 in the case of small amplitudes may result from the large dissipation by thermal conduction at higher temperature.  

\begin{figure*}
\epsscale{1.0}
\plotone{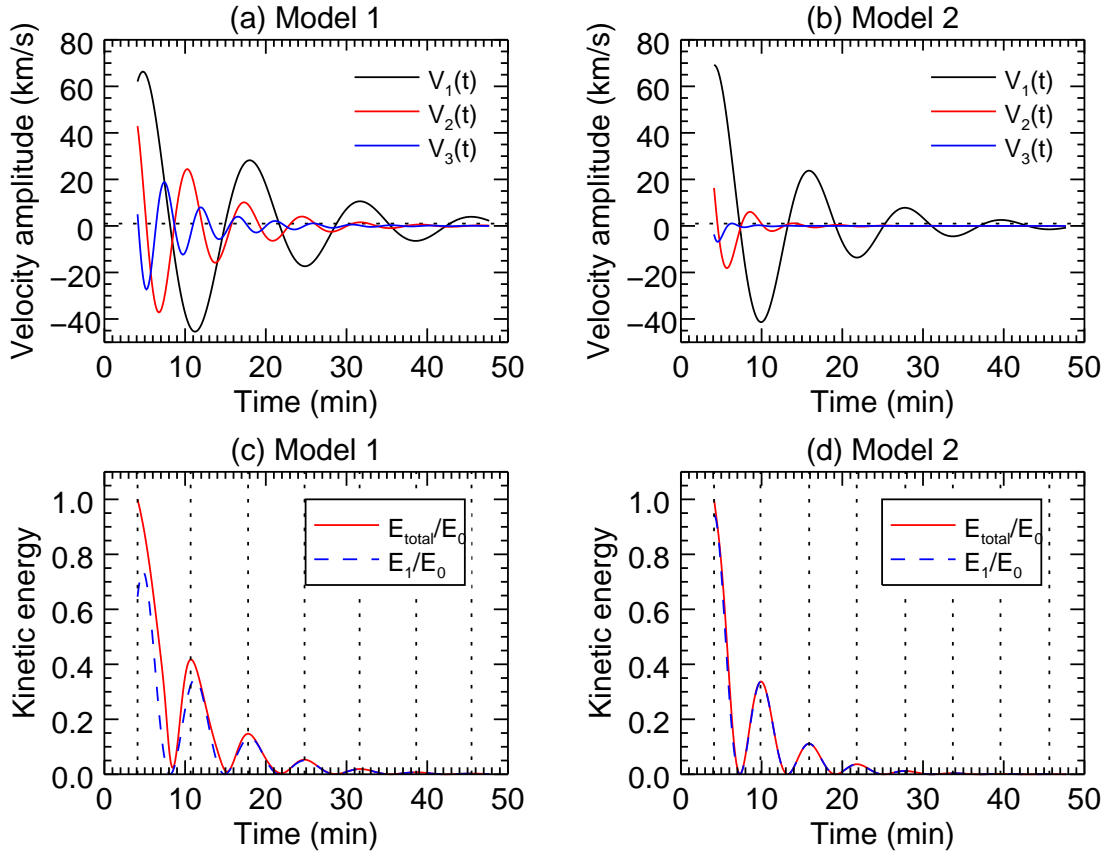}
\caption{ \label{fgfhw} (a) Time profiles of the amplitude of the Fourier components ($V_k(t\geqslant{t}_{\rm dur})$ with $k$=1, 2, and 3) for velocity oscillations along the loop for Model 1. (b) Same as (a) but for Model 2. (c) Time profiles of the total kinetic energy ($E_{total}$) of the waves normalized to its value ($E_0$) at $t=t_{dur}$ for Model 1. The dashed line represents the normalized kinetic energy ($E_1/E_0$) of the fundamental mode component. (d) Same as (c) but for Model 2.  }
\end{figure*}

In the second method, we directly analyze the simulations presented in Section~\ref{sctcom} using Fourier decomposition. As velocities along the loop satisfy the condition, $V(0, t)=V(L, t)=0$, after the flow driving at $x$=0 is stopped (i.e. $t>t_{\rm dur}$), the velocity profile $V(x, t)$ in $0\leqslant{x}\leqslant{L}$ at time $t$ can be extrapolated as an odd function into the domain $-L\leqslant{x}\leqslant{L}$. Then we can decompose $V(x, t)$ in Fourier sinus series in the $x$ direction as
\begin{equation}
 V(x, t)=\sum\limits_{k=1}^{\infty} V_k(t)\,{\rm sin}\left(\frac{\pi{kx}}{L}\right),
 \label{eqvxt}
\end{equation}
where
\begin{equation}
V_k(t)=\frac{2}{L}\int_0^L V(x ,t)\,{\rm sin}\left(\frac{\pi{kx}}{L}\right)dx.
\end{equation}
We calculate the amplitude $V_k(t)$ of the Fourier components for $k$=1--9. Figures~\ref{fgfhw}(a) and (b) illustrate the comparison between Model 1 and Model 2 in the temporal evolution of $V_k(t)$, indicating clearly that their fundamental mode components have a similar damping rate while the higher harmonics in Model 2 are damped much quicker than in Model 1. For a quantitative comparison, we measure the wave period and damping time of the decomposed components by fitting the amplitude profile $V_k(t)$ to an exponentially damped sine function. Figure~\ref{fgtdp}(b) shows the measured damping times with periods for the two models. It is noted that for Model 2 only three data points (corresponding to $k$=1--3) are available. No measurements for the higher harmonic components with $k\geqslant4$ are available because these harmonics are damped out within one wave period. We find that the first three Fourier components follow the scaling $\tau\propto{P}$ for Model 1 while they follow $\tau\propto{P^2}$ for Model 2, consistent with the results obtained using the first method. We also notice the difference from the first method in Model 1 that for the higher harmonic components with $k\geqslant4$ the damping time does not decrease with the wave period but varies in the range of about 6--11 minutes. This flattening feature could be caused by nonlinear mode coupling, through which the lower harmonics of large amplitudes leak energy into the high harmonics.

\begin{figure}
\epsscale{1.0}
\plotone{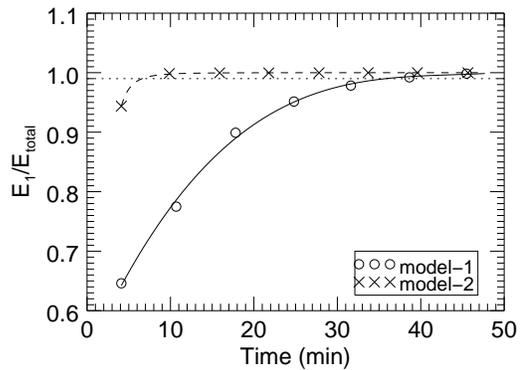}
\caption{ \label{fgftt} Ratio of the kinetic energy of the fundamental mode component ($E_1$) in the Fourier decomposition to the total kinetic energy ($E_{\rm total}$). The circles represent the values measured at the peak times of $E_{\rm total}$ for Model 1, and the crosses represent those for Model 2. The solid and dashed lines are the best fits (see the text). The horizontal dotted line indicates the ratio level of 0.99.  }
\end{figure}

To quantitatively estimate the excitation time of the simulated standing waves, we calculate the proportion of the kinetic energy of the fundamental mode component ($E_1(t)$) in the total kinetic energy of the waves ($E_{\rm total}(t)$). By defining the kinetic energy density as $\varepsilon(x, t)=V^2(x, t)/L$ and applying Equation~(\ref{eqvxt}), we obtain the total kinetic energy in the loop,
\begin{equation}
E_{\rm total}(t)=\int_0^L\varepsilon(x, t)\, dx=\frac{1}{2}\sum\limits_{k=1}^{\infty}V_k^2(t),
\end{equation}
and the kinetic energy of the Fourier $k$-harmonic component,
\begin{equation}
 E_k(t)=\int_0^L \frac{\left( V_k(t)\,{\rm sin}(\pi{kx}/L)\right)^2}{L}\, dx= \frac{1}{2}V_k^2(t).
\end{equation}
Thus it follows that $E_{\rm total}(t)=\sum_{k=1}^{\infty}E_k(t)$. Figures~\ref{fgfhw}(c) and (d) show the temporal evolution of $E_{\rm total}(t)$ and $E_1(t)$ calculated for the two models. Figure~\ref{fgftt} shows the ratios of $E_1(t)$ to $E_{\rm total}(t)$ calculated at the peak times of $E_{\rm total}(t)$, indicating that the proportion $E_1/E_{\rm total}$ tends to 1 in Model 2 much faster than in Model 1. We fit the data for Model 1 to a 3-degree polynomial $f(t)=a_0+a_1{t}+a_2{t^2}+a_3{t^3}$, and obtain $a_0$=0.53, $a_1$=0.031, $a_2$=$-7.0\times10^{-4}$, and $a_3$=$5.3\times10^{-6}$. We fit the data for Model 2 to a function in the form $f(t)=1-b_1\,{\rm ln}(1+b_2/t^{b_3})$ using the IDL function {\it curvefit}, and obtain $b_1$=3.1, $b_2$=3.1, and $b_3$=3.6. The main reason for fitting Model 2 to this functional form is to ensure the physical restriction $E_1/E_{\rm total}(t)\leqslant1$ to be met, and its quality of fit (with $\chi^2=1.8\times10^{-7}$) is much better than the polynomial fits (with $\chi^2=3.3\times10^{-4}$).

If assuming that a standing wave is set up when $E_1/E_{\rm total}\geqslant0.99$, we estimate the excitation time of a standing wave to be $t_{\rm exc}$=35.8 minutes for Model 1, while $t_{\rm exc}$=6.6 minutes for Model 2. We have measured the wave period of velocity oscillations, $P$=14.5 minutes for Model 1 and $P$=12.1 minutes for Model 2 (see Figures~\ref{fgtpf}(a) and (d)). Thus we get $t_{\rm exc}/P$=2.5 for Model 1 and $t_{\rm exc}/P$=0.5 Model 2. That is, for a velocity pulse as the wave exciter, the excitation of standing waves for Model 1 takes the time over about five reflections, while it takes only one reflection for Model 2. This confirms the results of mode identification based on the qualitative analysis in Section~\ref{sctcom}.

\section{Discussion}
\label{sctdis}

\subsection{Wave trigger mechanism}
We analyzed the magnetic configuration and related loop heating for a slow-mode wave event triggered by a footpoint flare. The NLFFF extrapolation and emission features such as circular ribbons with a remote brightening suggest that the wave event may be generated by slipping-type reconnections at a coronal null point in a fan-spine magnetic topology. We estimated the propagation speed of heat flux from the 3D loop length and time lag between the 1600 \AA\ light curves measured at two footpoints, and found that it is much faster than the wave propagation speed. This suggests that the spine loop may have been heated (to $\sim$10 MK) by energetic particles or heat flux from the reconnection region before the waves travel along it. Thus it is plausible to simulate the wave excitation in a hot loop. In addition, from the 1600 \AA\ light curve of the flare we estimated the impulsive heating time ($t_{\rm dur}\sim$4 minutes) and used it to constrain the duration of the wave driver for simulations.

The coordinated {\it Yohkoh}/SXT and {\it RHESSI} observations have shown that standing slow-mode waves in hot loops observed by SUMER were often associated with a footpoint brightening \citep{wan11}. As known in the literature, nearly all impulsively-generated slow-mode wave events observed with AIA occurred in a hot coronal loop heated by a confined flare at one footpoint displaying the feature of circular-like ribbons \citep[e.g.,][]{kum13, kum15, man16}, suggesting that they may be associated with a fan-spine topology like the case studied here. Recently, \citet{pan17} reported a standing slow-mode wave event suggesting a different trigger mechanism. The waves were triggered along coronal fan-like loops due to impact by a global EUV wave originating from a distant active region. The new case has a distinct feature that those oscillating loops are not involved in heating by flares. Whereas in our studied case the energy release process by the null-point reconnection is largely confined in a closed fan-spine field configuration, where longitudinal wave disturbances are trapped in the hot spine loop, which forms by filling hot plasma through chromospheric evaporation. A detailed explanation for flare trigger and related loop thermal dynamics can be found in \citet{sun13}. 

By analyzing SUMER spectra, \citet{wan05} found that the initiation of oscillations in hot loops is often associated with high-speed (100--300 km~s$^{-1}$) flow pulses, which may be produced by ejections of a small flux rope or mini-filament in fan-spine topology as shown in 3D simulations \citep[e.g.,][]{jia13, wyp17}. Signatures of related mini-filament eruptions were observed in the AIA 171 \AA\ and 304 \AA\ bands in some events \citep{kum13, kum15, man16}. Some 3D MHD simulations have shown that quasi-periodic outflows or single flow pulse injected at the footpoints of coronal loops inevitably generate slow magnetoacoustic waves propagating upwards along the loop \citep{ofm12, wan13, pro18}. 

Motivated by the above facts, we simulated the excitation of the observed wave event using a flow pulse injected at a footpoint using 1D models. As no traceable ejection was observed to be associated with the hot loop showing the longitudinal oscillations, we estimated the flow velocity from the maximum amplitude of density perturbations based on the linearized continuity equation. The measurements show that the initial perturbations (before reflection at the remote footpoint) simulated in the two cases (Model 1 and Model 2) both propagate at a speed close to the speed of sound, confirming they are the propagating slow magnetosonic waves.

\subsection{Validation of linear theory-based predictions}
\label{sctvwt}

We simulated the excitation of standing slow-mode waves in a hot loop observed with AIA using a 1D nonlinear MHD model with a flow driver at one footpoint. We compared the simulations in the two cases: (1) using the classical thermal conductivity and classical compressive viscosity (Model 1), and (2) using no thermal conduction but 15 times enhanced viscosity as determined using seismology technique based on linear MHD theory (Model 2).  We find that Model 2 can well reproduce several properties of the observations but Model 1 cannot. First, Model 2 can produce the standing wave pattern on a timescale consistent with the observed one, i.e., it takes only one reflection of the initial perturbation, while it needs about 5 reflections for Model 1. Second, Model 2 predicts the in-phase relationship between density and temperature perturbations in agreement with the observation, whereas Model 1 shows a large phase shift ($\sim60{\degr}$) between them. Third, Model 2 shows that the polytropic index, determined from the density and temperature scaling, is consistent with observations, i.e., its value lies close to $5/3$. In contrast, Model 1 shows a distinctly different value (1.390$\pm$0.003). Fourth, the wave period measured from the density perturbation for Model 2 is closer to the observational value than that for Model 1 (see the explanation given later on). The temporal profiles of density and temperature perturbations for Model 2 are harmonic and linear, which are more close to the observed features, while they show some nonlinearity for Model 1. The behavior of the two models is distinguishable by comparing Figures~\ref{fglcv}(c) and (d) with Figures~\ref{fgtpf}(b) and (e). The recovery of the observed wave period and damping time by Model 2 is expected because the used transport coefficients in this model are derived from the observational measurements based on the linear theory. The success of Model 2 thus validates the seismology technique we developed in \citetalias{wan15}. However, the result that Model 2 can recover the quick excitation of observed standing waves is unexpected, and the role of the transport coefficients in affecting the standing wave formation was not investigated before. We will discuss the wave excitation mechanism in the next section. In addition, as control experiments we have performed simulations for the two types of models with initial flow pulse of small amplitude of 0.01 $V_0$ (Models 1A and 2A; see Figures~\ref{fgmd1a} and~\ref{fgmd2a} in Appendix~\ref{sctapp}) or of short-duration of 1/2 $t_{\rm dur}$ (Models 1B and 2B; see Figures~\ref{fgmd1b} and~\ref{fgmd2b} in Appendix~\ref{sctapp}). These tests indicate that the main results obtained for Model 1 and Model 2 are robust, nearly independent of variations of the wave driver in amplitude and duration (see Table~\ref{tabmdl}).    

The result that the wave period in Model 2 is slightly shorter than that in Model 1 but more close to the observations can be explained in this way. In Model 2 as the thermal conduction is completely suppressed, the wave propagates with the faster adiabatic sound speed ($C_s$), while in Model 1 the wave propagates with a phase speed that is equivalent to the ``polytropic sound speed" ($C_p$) due to the energy loss by thermal conduction. For a polytropic process with $p=K\rho^\alpha$ where $K$ is a constant, it follows from the wave equation that
\begin{equation}
 C_p=\left(\frac{\alpha p_0}{\rho_0}\right)^{1/2}=\left(\frac{\alpha}{\gamma}\right)^{1/2}C_s.
\label{eqcp} 
\end{equation}
For $C_s=498$ km~s$^{-1}$ and $\alpha$=1.39 measured in Model 1, we obtain $C_p$=455 km~s$^{-1}$ which is consistent with the propagation speed of waves directly measured from the ridges in Figure~\ref{fgvp}. We also find that the predicted wave period, $P_{\rm pre}=2L/C_p$=13.2 minutes, agrees with that ($P$=13.5 minutes) measured from the density perturbation in Model 1.  For Model 2 we estimate $P_{\rm pre}=2L/C_s$=12.0 minutes, agreeing with the measured period ($P$=11.9 minutes), too. 

Comparing Model 1 with Model 2, we notice that their damping times are comparable. This makes it difficult to determine which is the dominant damping mechanism (thermal conduction or compressive viscosity) without referring to other properties such as the phase shift ($\Delta{\phi}$) between $T_1$ and $n_1$ and the polytropic index. In theory we may determine $\alpha$ from $C_p=2L/P$ using Equation~(\ref{eqcp}), and then use the relation between $\alpha$ and $\Delta{\phi}$ to estimate the effect of thermal conduction (see Equations~(\ref{eqphi}) and (\ref{eqalp})). However, since $C_p$ differs from $C_s$ by maximum a factor of $\gamma^{1/2}$ or $\sim$30\% (for $1\leqslant\alpha\leqslant\gamma$) and the accurate measurement of the 3D loop geometry is typically difficult, this method may not be applicable to observations. Thus the better way to constrain the effect of thermal conduction on the wave damping in observations is to measure $\Delta{\phi}$ as in \citetalias{wan15}. The following equations for $\Delta{\phi}$ can be derived using the 1D linear MHD theory when considering that thermal conduction dominates in the energy equation \citep[see][]{van11,wan15},
\begin{eqnarray}
  {\rm tan}\,\Delta{\phi}&=& 2\pi\gamma d,      \label{eqphi} \\
  (\gamma-1) {\rm cos}\,\Delta{\phi}&=& \alpha-1, \label{eqalp}
\end{eqnarray}
where $d$ is the thermal ratio \citep[see][]{dem03}, given by 
\begin{equation}
d=\frac{(\gamma-1)\kappa_\|T_0 \rho_0}{{\gamma}^2p_0^2 P}\approx 4.1\left(\frac{T_0^{3/2}}{n_0 P}\right),
\label{eqd}
\end{equation}
where $\rho_0=n_0 m_p$, $p_0=2n_0k_BT_0$, and $P=2\pi/\omega$ is the wave period. 

We validate the above phase shift relations based on Model 1. From the measured polytropic index $\alpha$=1.39, we obtain $\Delta{\phi}=54{\degr}$ using Equation~(\ref{eqalp}), which is consistent with the directly measured phase shift ($\Delta{\phi}=57{\degr}$) from the simulations. For Model 1 with $P$=13.5 minutes and $\tau$=11.6 minutes, we estimate $d$=0.052 and $\Delta{\phi}\approx29{\degr}$ using Equations~(\ref{eqphi}) and (\ref{eqd}). We find that the predicted phase shift by the linear theory is smaller than that from the simulation by about 50\%. To account for this difference, we recheck the derivation of Equations~(\ref{eqphi}) and (\ref{eqalp}) and notice that the approximation for the phase speed $V_p=\omega/k\approx C_s$ and neglecting the damping effect ({\it i.e.}, assuming $P/\tau\ll$1) were used in derivation. In general, given that the wave frequency ($\omega=\omega_r + i\omega_i$) is complex and $V_p=\omega_r/k$, we can obtain based on 1D linear wave theory
\begin{eqnarray}
{\rm tan}\,\Delta{\phi}&=&\frac{2\pi\gamma{d}\left(\frac{C_s}{V_p}\right)^2{\rm cos}\,\psi/\sqrt{1+\chi^2}}{1-2\pi\gamma{d}\left(\frac{C_s}{V_p}\right)^2{\rm sin}\,\psi/\sqrt{1+\chi^2}},  \label{eqphg} \\
(\gamma-1) {\rm cos}\,\Delta{\phi}&=& (\alpha-1) \left(1-\frac{2\pi\gamma{d}(C_s/V_p)^2{\rm sin\,\psi}}{\sqrt{1+\chi^2}}\right), \label{eqalg}
\end{eqnarray}
where $\chi=\omega_i/\omega_r$ and $\psi={\rm tan}^{-1}(\omega_i/\omega_r)$. It is obvious that the above equations will reduce to Equations~(\ref{eqphi}) and (\ref{eqalp}) on the condition of $V_p=C_s$ and $\omega_i$=0. 

We estimate the predicted phase shift using the improved equations in the two cases: (1) With $\omega_i$=0 and $V_p=\omega/k=2L/P$=444 km~s$^{-1}$, we obtain $\Delta{\phi}={\rm tan}^{-1}(2\pi\gamma{d}(C_s/V_p)^2)=35{\degr}$. (2) In the general case the values of $\omega_r$ and $\omega_i$ can be theoretically calculated from the dispersion relation for the fundamental standing wave (with $k=\pi/L$) using a normal mode analysis \citep[e.g.,][]{dem03}. Here by taking $\omega_r=2\pi/P$ and $\omega_i=1/\tau$ with $V_p=2L/P$, we obtain $\Delta\phi=37{\degr}$. We find that after correcting the error due to the assumptions in Equation~(\ref{eqphi}) the predicted value of $\Delta\phi$ is still much smaller (by about 35\%) than that from the simulation. This suggests that the underestimation may be caused by the nonlinear effect which needs further investigations in the future. 

\begin{deluxetable*}{llcccccccccc}
 \tabletypesize{\scriptsize}
 \tablecaption{Physical Parameters and Wave Properties for Different Models
 \label{tabmdl}}
 \tablewidth{0pt}
 \tablehead{
 \colhead{Model type} & \colhead{Model} & \colhead{$V_0$} & \colhead{$t_{\rm dur}$} & \colhead{$t_{\rm exc}$} & \colhead{$V_p^{\rm fit}$} & \colhead{$P_V$} & \colhead{$\tau_V$} & \colhead{$P_n$} & \colhead{$\tau_n$} & \colhead{$\Delta{\phi}$} & \colhead{$\alpha$} \\
 \colhead{} & \colhead{} & \colhead{($\rm km~s^{-1}$)} & \colhead{(min)} & \colhead{(min)} & \colhead{($\rm km~s^{-1}$)} & \colhead{(min)} & \colhead{(min)} & \colhead{(min)} & \colhead{(min)} & \colhead{} & \colhead{}
}
\startdata
  type I: & Model 1  & 115 & 4 & 35.8 & 504 & 14.5 & 11.2 & 13.5 & 11.6 & 57${\degr}$ & 1.390 \\
 $\kappa_\|=\kappa_\|^{\rm class}$,  & Model-1A & 1.15 & 4 & 37.2 & 401 & 15.1 & 10.0 & 14.1 & 11.8 & 51${\degr}$ & 1.432 \\
 $\eta=\eta_0^{\rm class}$ & Model 1B & 115 & 2 & 41.2 & 455 & 14.9 & 10.0 & 14.4 & 11.6 & 54${\degr}$ & 1.376 \\
\hline
  type II: & Model 2 & 115 & 4 & 6.6 & 619 & 12.1 & 10.6 & 11.9 & 12.2 & 0${\degr}$ & 1.684 \\
 $\kappa_\|=0$, & Model 2A & 1.15 & 4 & 7.4 & 585 & 12.4 & 9.5 & 12.2 & 12.8 & 0${\degr}$ & 1.667 \\
 $\eta=15\,\eta_0^{\rm class}$ &  Model 2B & 115 & 2 & 6.3 & 569 & 12.3 & 10.0 & 12.0 & 11.1 & 0${\degr}$ & 1.674 \\
\enddata
\tablecomments{Column 1 is the model type. Column 2 is the model name. Column 3 is the amplitude of the initial flow pulse. Column 4 is the pulse duration. Column 5 is the measured excitation time for a standing wave. Column 6 is the measured slope value of the first ridge in a time distance map of perturbed density. Columns 7 and 8 are the wave period and damping time, measured from velocity oscillations. Columns 9 and 10 are the same as columns 6 and 7 but for density oscillations. Column 11 is the measured phase shift between density and temperature oscillations. Column 12 is the measured polytropic index.}
\end{deluxetable*}

\subsection{Excitation mechanism of the fundamental standing mode}
\label{sctwem}

We estimated the excitation time of standing slow-mode waves from simulations using both the qualitative analysis based on the spatial and temporal features of the waves and the quantitative analysis based on the Fourier decomposition. We found that Model 2 with the anomalously large compressive viscosity and suppressed thermal conduction can excite the fundamental standing wave in a hot loop on a timescale well matching to the observation. Our control experiments show that this result is affected little by variations of the wave driver in amplitude and duration (see Appendix~\ref{sctapp}), providing additional support to the conclusion. It is noticed that numerical simulations with thermal conduction but no viscosity show the waves with strong nonlinearity which are obviously inconsistent with the observation as studied here \citep[e.g.,][]{men04, sig07, fan15}. This suggests the important role of viscosity in suppressing the nonlinear effect or smoothing high-frequency components in the waves. We analyzed the dependence of damping rate on wave frequency for the harmonic waves using two different methods: one by simulating each harmonic mode based on the initial velocity profile along the loop, and the other by decomposing the waves generated by the footpoint-driven flow pulse. The methods reveal a scaling law of $\tau\propto{P}$ for Model 1, while $\tau\propto{P^2}$ for Model 2. Considering that the damping times of the fundamental mode component for Model 1 and Model 2 are comparable, this implies that the ratio of their damping times for the harmonic number $k$ is $\tau_k^{\rm Model-1}/\tau_k^{\rm Model-2}\approx{k}$. That is, the $k$-harmonic component in Model 2 is damped $k$ times as quickly as that in Model 1. This explains why the fundamental mode can be set up in a much shorter time in Model 2. It is known that the linear slow wave theory predicts the scaling $\tau\propto{P^2}$ under the small dissipation assumption for viscosity \citep{por94, ofm00}. Our simulations here indicate that this scaling relation holds also in the regime of large viscosity (with the enhancement by more than an order of magnitude compared to the classical value). This property of viscosity is distinctly different from that of thermal conduction, whose effect on the wave damping becomes inefficient when the thermal conduction is very large due to the transition from adiabatic to isothermal behavior \citep{por94, dem03}.

The statistical studies of hot loop oscillations based on observations with {\it SOHO}/SUMER \citep{wan03a, wan07} and {\it Yohkoh}/BCS \citep{mar06} showed that these oscillations are best interpreted as the fundamental standing slow-mode waves \citep{wan11}. The spectral features of SUMER data suggested that these oscillations are triggered by hot flow pulses from one of the loop's footpoints, and the standing modes are often formed within one oscillation period \citep{wan05}. These properties appear to support the case of Model 2 (i.e., anomalously enhanced viscosity) as the dominant wave damping mechanism. However, an approximate linear scaling between damping time and wave period was found from observed oscillations, e.g., $\tau=0.68P^{1.06\pm0.18}$ obtained by fitting 49 cases in \citet{wan03a}, or a similar scaling obtained by fitting 35 cases in \citet{ofm02}. Their results appear to favor Model 1, i.e., thermal conduction as the dominant damping mechanism, which was first proposed by \citet{ofm02}. We suggest the following scenario to explain this paradox. It is known that for the dissipation of slow-mode waves by either ion viscous damping or electron conduction damping, the following relation can be derived from linear theory \citep{por94}
\begin{equation}
\tau\sim{C}(n_0/T_0^{3/2})P^2,
\label{eqtdp} 
\end{equation}
where $C$ is a constant. It implies that the scaling $\tau\propto{P^2}$ is valid only for a single loop or loops with the same temperature and density. However, this is not the case for a large number of samples in observation. For example, the SUMER observations showed that the loop temperatures are typically in the range 6--10 MK and the densities in the range $10^9$--$10^{10}$ cm$^{-3}$. If assuming hot loops follow the RTV scaling law \citep[$T_0\sim1.4\times10^3(pL)^{1/3}$;][]{ros78} and considering $P\sim2L/C_s$ for the fundamental mode, we can derive the relation $n_0/T_0^{3/2}\propto{1/P}$ from the RTV law by eliminating $L$. Thus from Equation~(\ref{eqtdp}) we find the scaling $\tau\propto{P}$ which is valid independently of the temperature and density of the sampled loops. This may explain the approximate linear scaling obtained from the SUMER observations. Note that the dispersion of data points to the fitted line is large \citep[see Figure~15 in][]{wan05}, indicating that many loops are not consistent with the predictions of static loop models. The above debates suggest that the observed $\tau$-$P$ scaling may not provide a tight constraint (or is not a sufficient condition) to determine or exclude whether the anomalously enhanced viscosity or the classical {\it Spitzer} conduction is the dominant wave dissipation mechanism. This implies the need for new statistical studies based on the AIA observations to verify whether the damping mechanism proposed to interpret the event studied here works only in this special case or in general.

The footpoint excitation of standing slow-mode waves in inhomogeneous loops (including the upper chromosphere, transition region, and gravitational stratification) was theoretically studied by \citet{tar05, tar07}. Their simulations with the effects of thermal conduction and radiation showed that the immediate excitation of the standing waves requires a special condition that the duration of heat pulse matches the period of the fundamental mode. However, this excitation condition is not supported by observations such as the wave event studied here and those observed with SUMER \citep{wan05}, which all showed that the heating duration is shorter than about the half wave period. Nevertheless the results of Model 2 presented in our study need to be validated based on similar inhomogeneous loop models in 1D or 2D in the future. 

\citet{wan15, wan16} have suggested that the suppression of thermal conduction in the event studied here is likely due to nonlocal conduction \citep{kar87}. The classical Spitzer form of conductivity is known to be valid under the assumptions that the electron velocity distribution is locally close to Maxwellian and the mean free path $\lambda$ is much smaller than the temperature gradient scale length $L_T$ \citep{ros86}. Such conditions may break down in solar flare loops with higher temperature because $\lambda$ increases with the squared temperature \citep[e.g.,][]{jia06, sha15}, resulting in the significant overestimation of heat flux \citep[the so-called saturation effect;][]{cow77, kar87, bat09}. For example, for hot loops with $T$=10 MK, if assuming $n=10^9$ cm$^{-3}$ in a large ($L$=100 Mm) loop or $n=10^{10}$ cm$^{-3}$ in a small ($L$=10 Mm) loop, we estimate that $\lambda/L_T\approx0.1$ using $\lambda/L_T= 0.1 (T /10\,{\rm MK})^2 [(L/100\,{\rm Mm})(n/10^9\,{\rm cm}^{-3})]^{-1}$ \citep{ros86}. This estimate suggests the breakdown of the diffusion approximation in Spitzer conduction theory that requires $\lambda/L_T\lesssim0.015$ \citep{gra80, ros86}. Some recent studies showed that turbulent magnetic fluctuations also can significantly reduce the parallel thermal conductivity in flaring coronal loops \citep{bia16a, bia16b, bia18}. In addition, the suppressed thermal conduction predicts a weaker chromospheric evaporation \citep{kar87}, and thus may imply a smaller-than-expected density in hot oscillating loops. This appears to be supported by the SUMER observations showing that except for initial flow pulses no persistent background flow was found in hot loops \citep{wan05}. The density deficit caused by the conduction suppression may be estimated based on the ${\rm EM}$-$T$ correlation for flare loops where $T$ is the peak temperature and ${\rm EM}$($\simeq{n^2L^3}$) the volume emission measure  \citep{fel95, shi99, shi02}. Assuming a balance between conduction cooling and reconnection heating and the pressure balance of flare loops, \citet{shi99} derived the scaling law ${\rm EM}\propto B^{-5}T^{17/2}$, where $B$ is the magnetic field strength. We define the suppressed conductivity as $\kappa_S = \kappa_0/S$, where $\kappa_0\simeq10^{-6}$ cgs is the thermal conductivity of Spitzer and $S$ the suppression factor. By considering a loop is heated to the same temperature in the two cases, i.e. with or without conduction suppression, we can obtain the modified scaling law ${\rm EM}_S\propto S^{-3}B^{-5}T^{17/2}$ and the density ratio $n_S/n=S^{-3/2}$.  Given $S\gtrsim$3 as measured in \citetalias{wan15}, for instance, we expect that the conduction suppression will lead to the flare loop underdense by at least a factor of 5. Thermal conduction suppression may provide an alternative explanation for the finding that AR hot loops tend to be underdense compared to the hydrostatic predictions \citep{win03, rea14}.

The reason for anomalous enhancement of compressive viscosity in the event studied here is unclear, but it is known that anomalous viscosity can be caused by a process such as thermal non-equilibrium between electrons and ions in the impulsively heated loops, which in this event is likely due to continuous heating by slow reconnection at quasi-separatrics layers (QSLs) and null-point \citep[e.g.,][]{sun13, qiu16, zhu18}. Turbulence is also a possible process that can lead to an enhanced viscosity such as Bohm diffusion \citep{boh49} and eddy viscosity \citep{hol88}.

\section{Conclusions}
\label{sctccs}
In conclusion, we have found that a standing slow-mode wave event was triggered by a flare in a closed fan-spine magnetic topology. The footpoint excitation of the wave event is simulated based on a 1D nonlinear MHD loop model for two sets of parameters. In one case with anomalously large compressive viscosity and suppressed thermal conduction, the standing wave pattern can be produced quickly on a timescale that is consistent with the observation, whereas in the other case with the classical conduction and viscosity the formation of the standing wave takes many wave reflections in the numerical model. In this case, basically a reflecting propagating wave is excited. By analyzing the dissipation properties of harmonic waves, we find that the scaling law between damping time and wave period follows $\tau\propto{P^2}$ in the former case while $\tau\propto{P}$ in the latter case. This implies a more efficient dissipation of the higher harmonic components when the viscosity is strongly enhanced, so explaining the quick formation of the fundamental standing waves. Whether this is a common excitation mechanism requires further validation by studying a large sample of SDO/AIA wave events using a similar method as that employed in \citetalias{wan15}.

\acknowledgments
The work of TW and LO was supported by the NASA Cooperative Agreement NNG11PL10A to CUA. TW also would like to thank Max-Planck-Institut f\"{u}r Sonnensystemforschung for support to a short-period visit research. The work of SKS has been partially supported by the BK21 plus program through the National Research Foundation (NRF) funded by the Ministry of Education of Korea. SDO is a mission for NASA’s Living With a Star (LWS) program.

\appendix

\section{Simulations of control experiments}
\label{sctapp}

\begin{figure*}
\epsscale{1.0}
\plotone{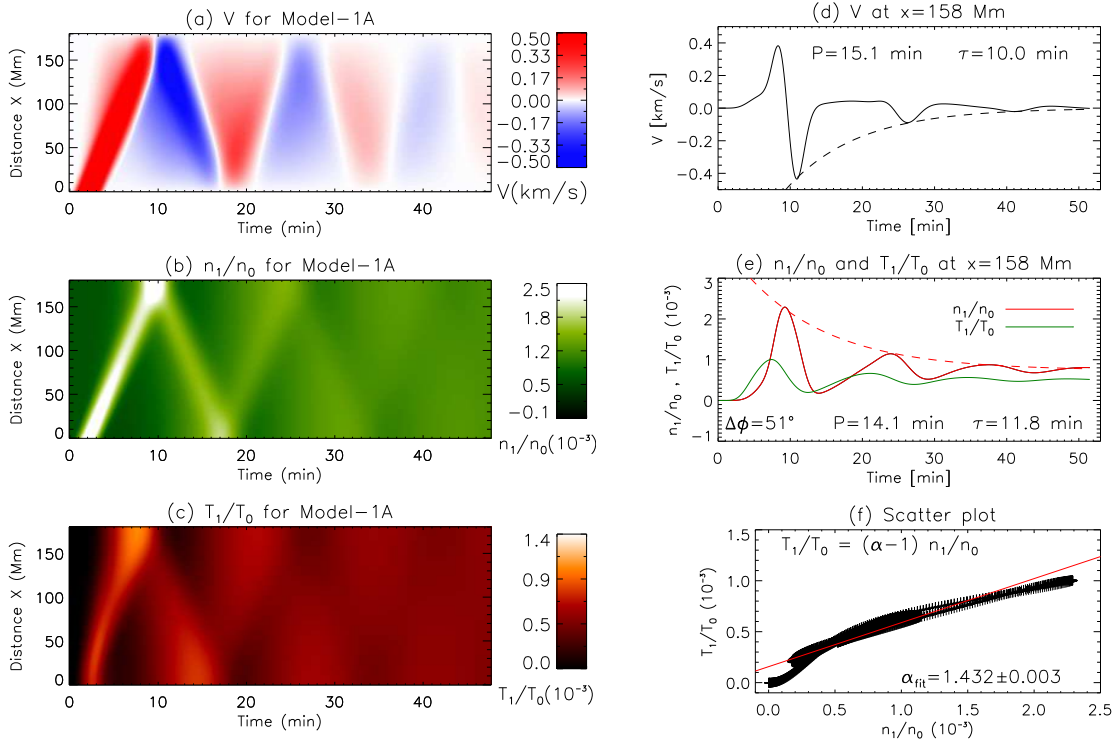}
\caption{ \label{fgmd1a} Simulations of a control experiment (Model 1A) that has the same parameters as Model 1 but with the initial pulse amplitude $V_0$=0.0023 $C_s$. (a)-(c) Time distance maps for velocity ($V$), perturbed density ($n_1/n_0$), and perturbed temperature ($T_1/T_0$) along the loop. Temporal profiles of (d) $V$, and (e) $n_1/n_0$ and $T_1/T_0$ at the location $x=158$ Mm. (f) The scatter plot (pluses) and its best fit (solid line). In (d)-(f) the measured oscillation period ($P$), decay time ($\tau$), phase shift ($\Delta{\phi}$) between $n_1$ and $T_1$, and polytropic index ($\alpha_{\rm fit}$) are marked on the plots.}
\end{figure*}

\begin{figure*}
\epsscale{1.0}
\plotone{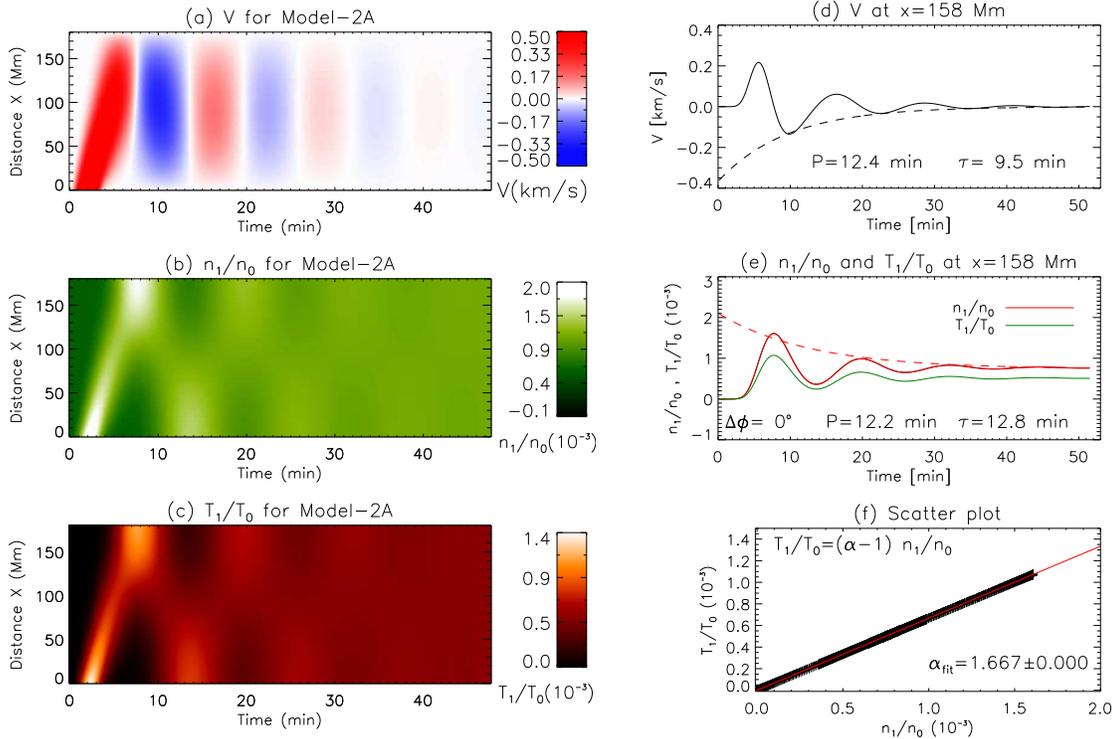}
\caption{ \label{fgmd2a} Simulations of a control experiment (Model 2A) that has the same parameters as Model 2 but with the initial pulse amplitude $V_0$=0.0023 $C_s$. The annotations are the same as in Figure~\ref{fgmd1a}. }
\end{figure*}

 With control experiments we examine whether differences between the behavior of Model 1 and Model 2, particularly in excitation time of a standing wave, is affected by variations in the wave driver. In the first case, we test the dependence of the model behavior on the amplitude of the initial flow pulse. We design two models, called Model 1A and Model 2A, which have the same physical parameters as Model 1 and Model 2, respectively, but have the pulse amplitude smaller by a factor of 100, i.e. taking $V_0$=1.15 km~s$^{-1}$. Figures~\ref{fgmd1a} and~\ref{fgmd2a} show the simulation results for these two models. We measure the wave period, damping time, polytropic index, and phase shift between density and temperature perturbations (see also Figure~\ref{fgntc}(a)). The comparison with those in Model 1 and Model 2 indicates that the dependence of these wave properties on the pulse amplitude is weak. Figure~\ref{fgprp}(a) shows the measurements of the phase speed for initial propagating waves. We find that the propagation speed in Model 1 is reduced by 20\% ($V_p^{\rm fit}$=401 km~s$^{-1}$ in Model 1A) for a small amplitude of the pulse, while the one in Model 2 is reduced by only $\sim$5\% ($V_p^{\rm fit}$=585 km~s$^{-1}$ in Model 2A). This indicates that the phase speed of initial perturbations in Model 2 is not apt to be affected by the amplitude variability of the driver compared to Model 1. The fact could be attributed to the suppression of nonlinearity by the enhanced viscosity in Model 2.  

\begin{figure*}
\epsscale{1.0}
\plotone{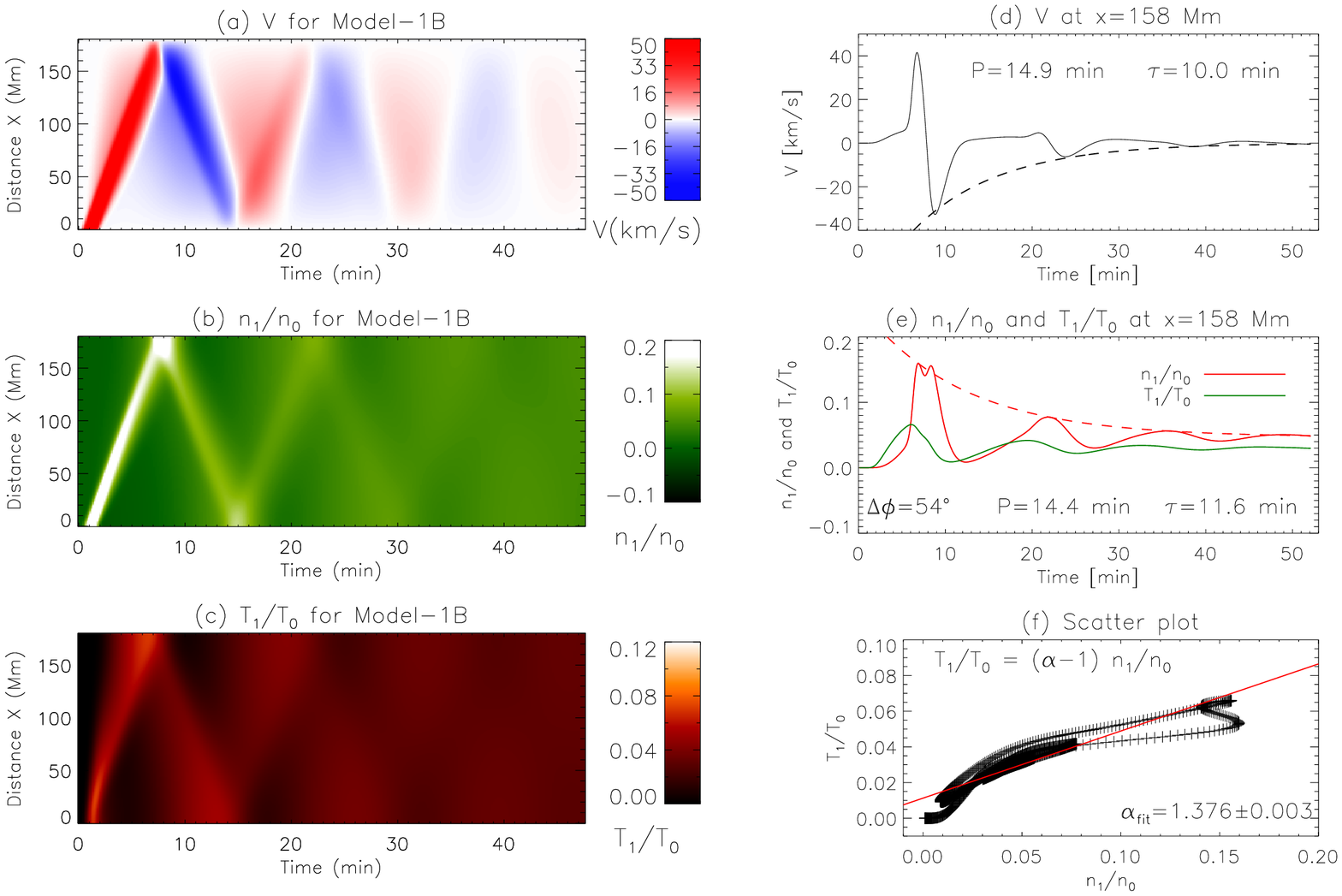}
\caption{ \label{fgmd1b} Simulations of a control experiment (Model 1B) that has the same parameters as Model 1 but with the initial pulse duration $t_{\rm dur}$=2 minutes. The annotations are the same as in Figure~\ref{fgmd1a}.}
\end{figure*}

\begin{figure*}
\epsscale{1.0}
\plotone{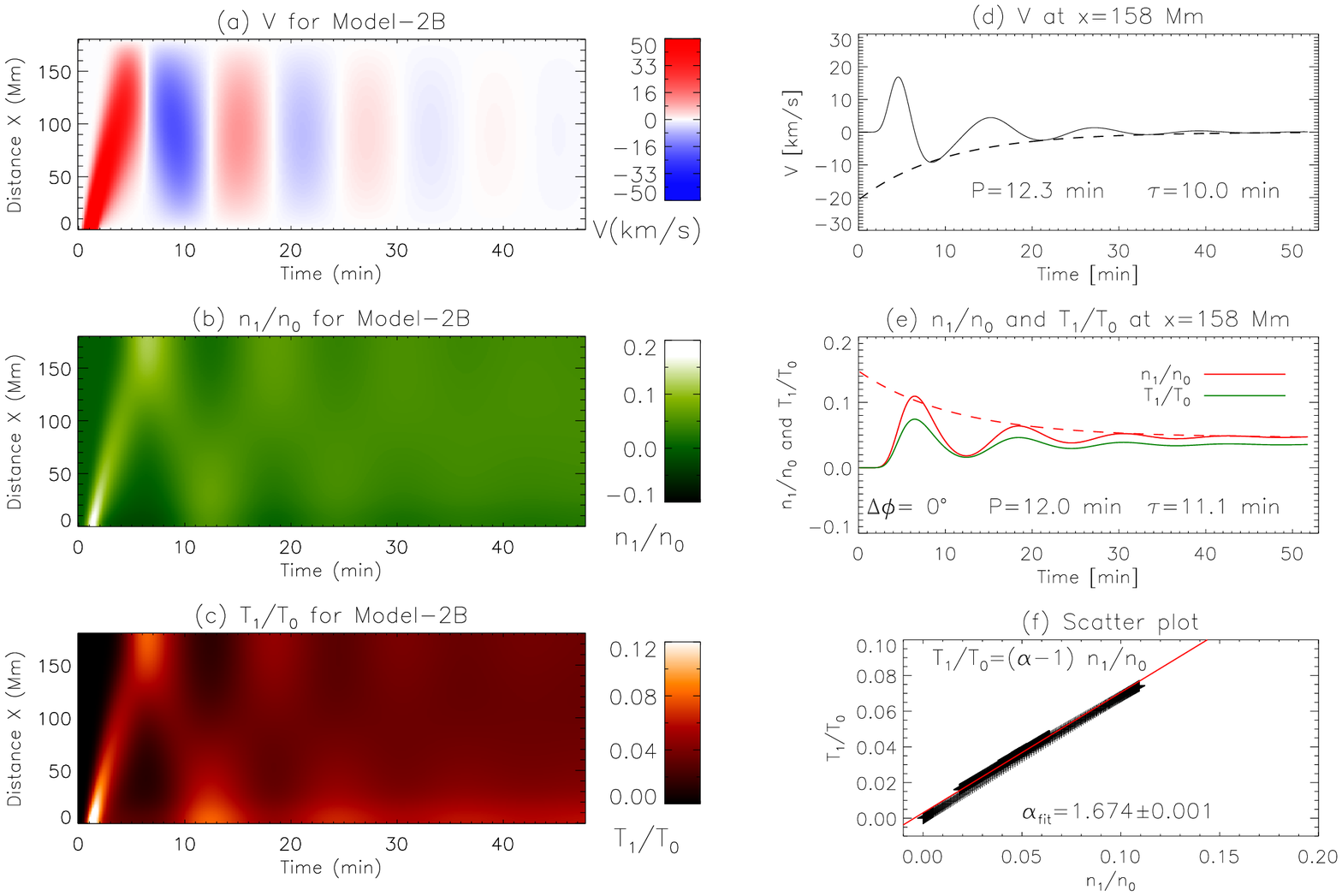}
\caption{ \label{fgmd2b}  Simulations of a control experiment (Model 2B) that has the same parameters as Model 2 but with the initial pulse duration $t_{\rm dur}$=2 minutes. The annotations are the same as in Figure~\ref{fgmd1a}.}
\end{figure*}

\begin{figure*}
\epsscale{1.0}
\plotone{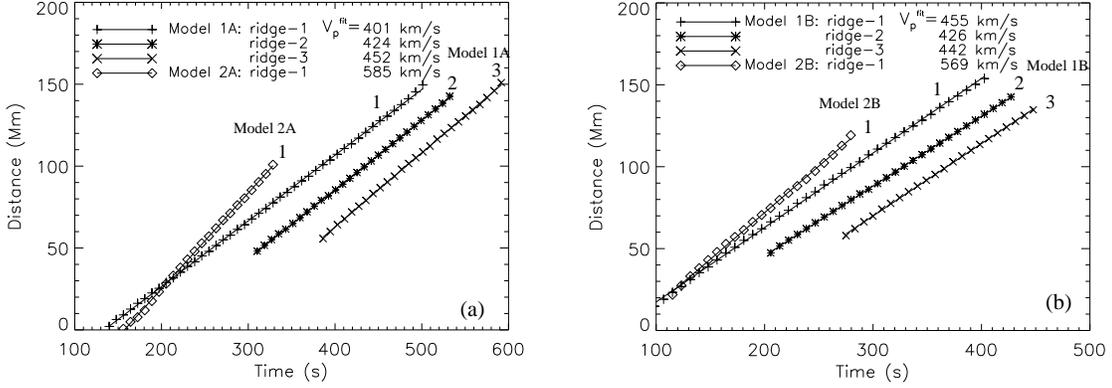}
\caption{ \label{fgprp} Measurements of the phase speed of propagating slow waves. (a) The peak position of density perturbations along the loop measured for three ridges of Model 1A: ridge-1 (pluses) during $t$=[140, 501] seconds, ridge-2 (asterisks) during $t$=[706, 928] seconds, and ridge-3 (crosses) during $t$=[1166, 1372] seconds.  For ridge-2 the peak positions are measured from $x=L$ and are plotted with the time of $t-396$ seconds. For ridge-3 the peak positions are plotted with the time $t-780$ seconds. The peak positions of ridge-1 (diamonds) for Model 2A are measured during $t$=[156, 328] seconds. The solid lines are the best fit to the data points, and their slope values ($V_p^{\rm fit}$) are marked on the plot. (b) For three ridges of Model 1B: ridge-1 (pluses) during $t$=[99, 403] seconds, ridge-2 (asterisks) during $t$=[584, 805] seconds, and ridge-3 (crosses) during $t$=[1043, 1216] seconds.  For ridge-2 the peak positions are measured from $x=L$ and are plotted with the time of $t-378$ seconds. For ridge-3 the peak positions are plotted with the time $t-768$ seconds. The peak positions of ridge-1 (diamonds) for Model 2B are measured during $t$=[115, 279] seconds. The lines have the same meaning as in (a). }
\end{figure*}

\begin{figure*}
\epsscale{1.0}
\plotone{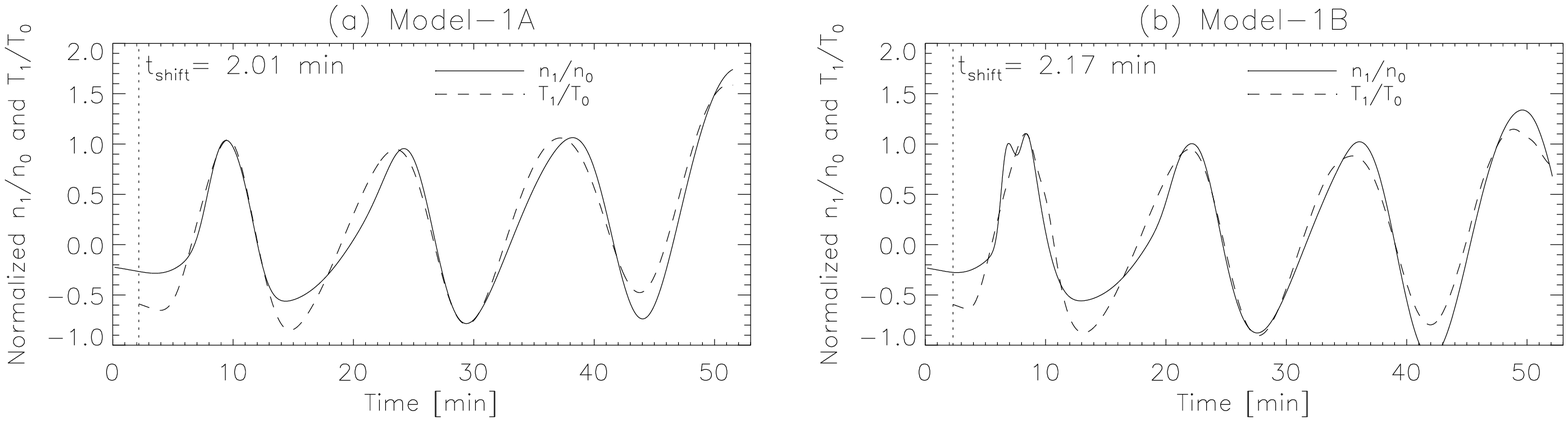}
\caption{ \label{fgntc} Measurements of the phase shift between the perturbed density and temperatures. (a) The amplitude-normalized density (solid line) and temperature (dashed line) perturbations for Model 1A. The temperature profile has been corrected relative to the density profile by a phase shift of $t_{\rm shift}$=2.01 minutes, which corresponds to the maximum correlation between them. (b) Same as (a) but for Model 1B, where the measured phase shift $t_{\rm shift}$=2.17 minutes. }
\end{figure*}

In the second case, we test the dependence of the model behavior on the duration of the initial pulse. We design Model 1B and Model 2B same as Model 1 and Model 2, respectively, but with the pulse duration shorter by a factor of 2, i.e. taking $t_{\rm dur}$=2 minutes. Figures~\ref{fgmd1b} and~\ref{fgmd2b} show the simulation results for these two models. The comparison with those for Model 1 and Model 2 indicates that the dependence of the wave properties on the pulse duration is weak. In a summary, we list the measured wave properties for the different models in Table~\ref{tabmdl}.

\begin{figure}
\epsscale{0.6}
\plotone{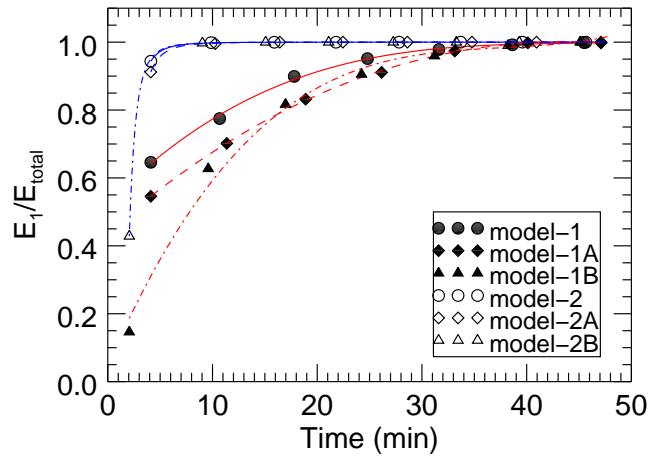}
\caption{\label{fgetf} Ratio of the kinetic energy of the fundamental mode component ($E_1$) in the Fourier decomposition to the total kinetic energy ($E_{\rm total}$), measured at the peak times of $E_{\rm total}$. Different symbols represent the data for the different models, whose best fits are indicated with the different lines: red solid line for Model 1, red dashed line for Model 1A, red dot-dashed line for Model 1B, blue solid line for Model 2, blue dashed line for Model 2A, and blue dot-dashed line for Model 2B.}
\end{figure}

Finally, we emphasize our conclusion that the different behavior of the two types of models (Model 1 and Model 2) is mainly due to their difference in transport coefficients, and particularly the anomalously large viscosity is crucial in leading to a quick formation of the fundamental standing wave in flaring loops. This conclusion is supported by the results of control numerical experiments. We find that the excitation time of the fundamental standing mode is nearly independent of the wave driver's amplitude and duration, as evidenced by spatial and temporal features of the waves in velocity and density (see Panels (a) and (b) in Figures~\ref{fgmd1a}--\ref{fgmd2b}). We also  quantitatively measure the excitation time of the fundamental mode for the control numerical experiments using the Fourier decomposition analysis (see Figure~\ref{fgetf}). We fit the data of $E_1/E_{\rm total}$ to a 3th-degree polynomial for Model 1 (A and B), and obtain the coefficients: $a_0$=0.44, $a_1$=0.027, $a_2$=$-3.5\times10^{-4}$, and $a_3$=$6.0\times10^{-7}$ for Model 1A, and $a_0$=0.047, $a_1$=0.071, $a_2$=$-1.9\times10^{-3}$, and $a_3$=$1.6\times10^{-5}$ for Model 1B. The numerical results for Model 2 (A and B) are fitted to a function in the form $f(t)=1-b_1\,{\rm ln}(1+b_2/t^{b_3})$. We obtain $b_1$=4.2, $b_2$=4.3, and $b_3$=3.8 for Model 2A, and $b_1$=3.0, $b_2$=3.0, and $b_3$=3.7 for Model 2B. We define the excitation time of the standing fundamental mode as the time when $E_1/E_{\rm total}\geqslant0.99$. The measurements confirm our conclusion (see Column 5 in Table~\ref{tabmdl}).

\end{document}